\documentclass[aps,pre,onecolumn]{revtex4}
\usepackage{amsmath}
\usepackage{epsfig}
\usepackage{titlesec}
\usepackage{amssymb,graphicx}
\usepackage{amsmath,bm}
\newfont{\logo}{logo10}

\usepackage[hang,nooneline]{subfigure}                         
\newcounter{fig}   
\usepackage{multirow}                                          
\usepackage{hhline}                                            

\newcommand{\bea}{\begin{eqnarray}}
\newcommand{\eea}{\end{eqnarray}}
\newcommand{\bes}{\begin{subequations}}
\newcommand{\ees}{\end{subequations}}

\begin{document}

\title{
Vector rogue waves and dark-bright boomeronic solitons in autonomous and nonautonomous settings}

\author{R. Babu Mareeswaran\footnote{Email: babu\_nld@rediffmail.com}}
\affiliation{Post Graduate and Research Department of Physics, Bishop Heber College, Tiruchirapalli--620 017, Tamil Nadu, India}
\author{E. G. Charalampidis\footnote{Email: charalamp@math.umass.edu}}
\affiliation{School of Civil Engineering, Faculty of Engineering, Aristotle University of Thessaloniki, Thessaloniki 54124, Greece\\}
\affiliation{Department of Mathematics and Statistics, University of Massachusetts, Amherst, MA 01003-4515, USA}
\author{T. Kanna\footnote{Email: kanna\_phy@bhc.edu.in}}
\affiliation{Post Graduate and Research Department of Physics, Bishop Heber College, Tiruchirapalli--620 017, Tamil Nadu, India\\}
\author{P. G. Kevrekidis\footnote{Email: kevrekid@math.umass.edu}}
\affiliation{Department of Mathematics and Statistics, University of Massachusetts, Amherst, MA 01003-4515, USA}
\author{D. J. Frantzeskakis\footnote{Email: dfrantz@phys.uoa.gr}}
\affiliation{Department of Physics, University of Athens, Panepistimiopolis, Zografos, Athens 15784, Greece}

\date{\today}

\begin{abstract}
In this work, we consider the dynamics of vector rogue waves
and dark-bright solitons in two-component nonlinear Schr{\"o}dinger
equations with various 
physically motivated
time-dependent nonlinearity coefficients,
as well as spatio-temporally dependent
potentials. A similarity transformation is utilized to convert
the system into the integrable Manakov system
and subsequently the vector rogue and dark-bright boomeron-like soliton
solutions of the latter are converted back into ones of the original non-autonomous model.
Using direct numerical simulations we find
that, in most cases, the rogue wave formation is rapidly followed by a modulational
instability that leads to the emergence of an expanding soliton train. Scenarios different than
this generic phenomenology are also reported.
\end{abstract}

\maketitle

\section{Introduction}

Over the past few years, models of atomic physics~\cite{emergent} and nonlinear optics~\cite{kiag}
with a spatially or temporally dependent nonlinearity have
received an increasing amount of attention.
More specifically, in the context of atomic Bose-Einstein condensates (BECs), the mean-field
Gross-Pitaevskii model has been examined in the presence
of temporally~\cite{BEC} or spatially~\cite{boris} varying
nonlinearity coefficients. These can be realized also experimentally in this
setting (by suitable tuning of the $s$-wave scattering length in space or time)
by employing external magnetic \cite{mFRM} or optical \cite{oFRM} fields close to Feshbach resonances.
On the side of nonlinear optics,
nonlinear Schr{\"o}dinger systems of variable (periodic) dispersion have been explored in
connection to dispersion management~\cite{tur} and the corresponding
notion of nonlinearity management was also theoretically
developed~\cite{towers,pelinov} and experimentally tested both
in the context of solitary waves and in that of prototypical instabilities,
such as the modulational instability (MI)~\cite{psaltis}.

On the other hand, one of the prototypical coherent structures
that have emerged in recent years as being of relevance to an
ever-expanding range of settings consists of rogue waves
(also known as freak or extreme waves)~\cite{k2}.
Such waveforms have recently been observed in various experiments carried out in
a diverse host of systems, including nonlinear optics \cite{opt1,opt2,opt3}, mode-locked lasers \cite{laser},
superfluid helium \cite{He}, hydrodynamics \cite{hydro}, Faraday surface ripples
\cite{fsr}, parametrically driven capillary waves \cite{cap}, and plasmas \cite{plasma}.
At the same time, there has been a large volume of
theoretical effort aimed at understanding the structural characteristics
and dynamical properties of such waves. A number of recent reviews
have attempted to summarize different aspects of this
effort~\cite{yan_rev,solli2,onorato}. Much of this theoretical
understanding has, interestingly, been focused 
on special solutions
of the focusing single-component nonlinear Schr{\"o}dinger (NLS) equation. These
solutions include the rational waveform, first proposed by Peregrine~\cite{pere},
and its generalizations, proposed by Kuznetsov \cite{kuz}, Ma \cite{ma},
and Akhmediev \cite{akh}, as well as Dysthe and Trulsen \cite{dt} among others.

On the other hand, more recently, there has been a number of studies devoted
to rogue waves in multi-component (vector) NLS equations. Generally, such NLS systems have been
studied extensively both in the physics of atomic BECs (where they describe different
atom species, or mixtures of different spin states of the same atom species \cite{emergent}) and
in nonlinear optics (where they describe the dynamics of two waves of different frequencies,
or two waves of different polarizations in a nonlinear birefringent medium \cite{kiag}).
Furthermore, two-component NLS equations describe two-wave systems in deep water
(crossing sea states), where the increase of the MI growth rate and
enlargement of the instability region, was proposed as a possible mechanism for the emergence
of extreme wave events \cite{onor2}. Rogue waves in such two-component NLS equations have been studied
in BEC mixtures \cite{konotop,pors} and optics, in connection with the four-wave
mixing process \cite{fwm}, but also in the context of other multi-wave
systems --such as the Maxwell-Bloch \cite{mb} and long-short-wave interaction \cite{swlw} systems.
From a mathematical point of view, exact analytical rogue wave solutions of the two coupled
NLS equations were presented for the completely integrable Manakov case \cite{manakov} in Ref.~\cite{degasPRL},
(see also the recent work~\cite{ML1})
while higher-order rogue waves pertaining to this case were recently studied as well \cite{horw}.

Our efforts in the present paper will be in blending these settings,
i.e., in obtaining rogue wave, as well as boomeron-like
solutions (that spontaneously reverse their direction --see Ref.~\cite{degasp}
and references therein) in two-coupled NLS equations with time-dependent
nonlinearities that could be of relevance, as indicated
above, both in atomic BECs, as well as --potentially-- in nonlinear optics.
The technique that we will use relies on
performing
suitable nonlinear transformations
to convert the original non-autonomous
model into an integrable one, for which such solutions (rogue waves or boomerons)
can be identified --see, e.g., Refs.~\cite{serkin1,jvv1,jvv2,laks,yan1,lihe1,xuhe3,hefokas2,kannajpa} and
references therein. However, as noted above, the starting point will be different
from these earlier works in that we will use a two-component
variant of the NLS equation, featuring both a nonlinearity prefactor
dependence on the evolution variable, as well as a parabolic potential
(of relevance to BECs) with a frequency dependence on the evolution variable; both linear
and nonlinear coupling between the two ``species'' will be present.
%
%
Starting from this non-autonomous, yet experimentally realizable
variant of the two-component model, suitable transformations
allow us to translate the dynamical evolution into the integrable
Manakov form. In the latter context, the work of Ref.~\cite{konotop}
enables us to examine rogue waves, while the findings of Ref.~\cite{degasPRL}
allow us to examine the interaction of the rogue wave
with a dark-bright ``boomeronic'' structure. Furthermore, we
examine the robustness of these structures in direct numerical simulations
of the original non-autonomous model. Here, generalizing the conclusions
of the recent work~\cite{He2014577} to multiple components, we find
that in most cases examined, the background state becomes unstable, giving rise to MI
and an expanding array of bright solitons.

Our presentation will be structured as follows. In Sec. II, we
present the theoretical formulation of the model, and
explore its realizability in current experimental setups.
In Sec. III, we turn to numerical results,
where we examine the dynamical evolution of the different solutions.
Finally, in Sec. IV, we summarize our findings and present some
conclusions, as well as future challenges.

\section{Theoretical Model and Analysis}
\subsection{The autonomous case}
We start by considering a system of two coupled NLS equations, in $(1+1)$ dimensions,
with constant nonlinear coefficients; this system is relevant
to both atomic BEC physics~\cite{emergent} and nonlinear optics~\cite{kiag} and
can be expressed in the following dimensionless form:
\bea
i E_{1,z}+E_{1,tt}+2\gamma \sum_{l=1}^2|E_l|^2E_1+\alpha E_1+\sigma E_2=0,\nonumber\\
i E_{2,z}+E_{2,tt}+2\gamma \sum_{l=1}^2|E_l|^2E_2-\alpha E_2+\sigma E_1=0,
\label{twogp}
\eea
where $E_1$ and $E_2$ are the complex electric field envelopes in the context of optics
or the wavefunctions of two distinct hyperfine states
(e.g., of $^{87}$Rb for $\gamma<0$, or of $^7$Li for $\gamma>0$) in the context of BECs.
Additionally, $z$ and $t$ are the longitudinal 
and transverse coordinates in the optical setting, while in BECs $z$ plays the role of
time as the evolution variable and $t$ is the spatial coordinate. The parameter $\gamma$
represents the nonlinear coefficient while the self-focusing and defocusing case (attractive
and repulsive interatomic interactions in BECs) corresponds to $\gamma>0$ and $\gamma<0$, respectively.
Finally, $\sigma$ is the normalized linear coupling constant induced by a periodic twist 
of the birefringence axes in optics (see, e.g., Ref.~\cite{sciencekanna}), or a Rabi coupling
in atomic BECs (see, e.g., Refs.~\cite{Rabi1,Rabi2}), while $\alpha$ is the phase-velocity 
mismatch from resonance, i.e., a frequency detuning parameter.

We now apply the following rotational transformation~\cite{sciencekanna,Rabi1,Rabi2} to 
Eq.~(\ref{twogp}) in the case of $\gamma$=$1$ ($\gamma$ will be assumed, hereafter, to be
positive, as we will consider the self-focusing and attractive interaction setting),
\begin{equation}
\begin{pmatrix}
E_{1} \\ E_{2}
\end{pmatrix}
=\begin{pmatrix}
\cos{\theta} & - \sin{\theta} \\
\sin{\theta} &   \cos{\theta}
\end{pmatrix}
\begin{pmatrix}
 q_{1}e^{i\Gamma z} \\
 q_{2}e^{-i\Gamma z}
 \end{pmatrix},
 \label{rabi}
\end{equation}
where $\Gamma = \sqrt{\alpha^2+\sigma^2}$ and $\theta = \frac{1}{2}\tan^{-1}(\frac{\sigma}{\alpha})$;
this way, the resulting equation becomes the self-focusing integrable Manakov system~\cite{manakov}
(see also~\cite{PRLkanna}):
\bea
iq_{j,z}+q_{j,tt}+2\sum_{\ell=1}^2 |q_\ell|^2q_j=0,\quad j=1,2.
\label{twomanakov}
\eea
An interesting rogue wave solution of the Manakov system (\ref{twomanakov}) has been obtained in
Ref.~\cite{degasPRL} using the Darboux transformation:
\bes\bea
q_1(t,z) =\left(\frac{Ga_1+Ma_2}{F}\right)e^{2i\omega z},\\
q_2(t,z) =\left(\frac{Ga_2-Ma_1}{F}\right)e^{2i\omega z},
\eea\ees
where $G = (3/2) - 8\omega^2z^2-2a^2t^2+8i\omega z+|f|^2e^{2at}$, $M = 4f(at-2i\omega z-\frac{1}{2})e^{at+i\omega z}$, $F = (1/2)+8\omega^2z^2+2a^2t^2+|f|^2e^{2at}$, $a = \sqrt{a_1^2+a_2^2}$, $\omega = a^2$, $a_1$ and $a_2$ are two arbitrary real parameters, and $f$ is a complex parameter.
It should be pointed out that the amplitude and intensity of the above
rogue wave are different in the two components unlike the rogue waves with
the same amplitude in both components~\cite{konotop}. Another striking feature 
of this solution is that the rogue waves co-exist with dark-bright (boomeron-like)
solitons. By using the transformation (\ref{rabi}), we construct exact
rogue wave solutions of the system (\ref{twogp}).

\begin{figure}[!pht]
\begin{center}
\vspace{-0.7cm}
\mbox{\hspace{-0.2cm}
\subfigure[][]{\hspace{-1.0cm}
\includegraphics[height=.21\textheight, angle =0]{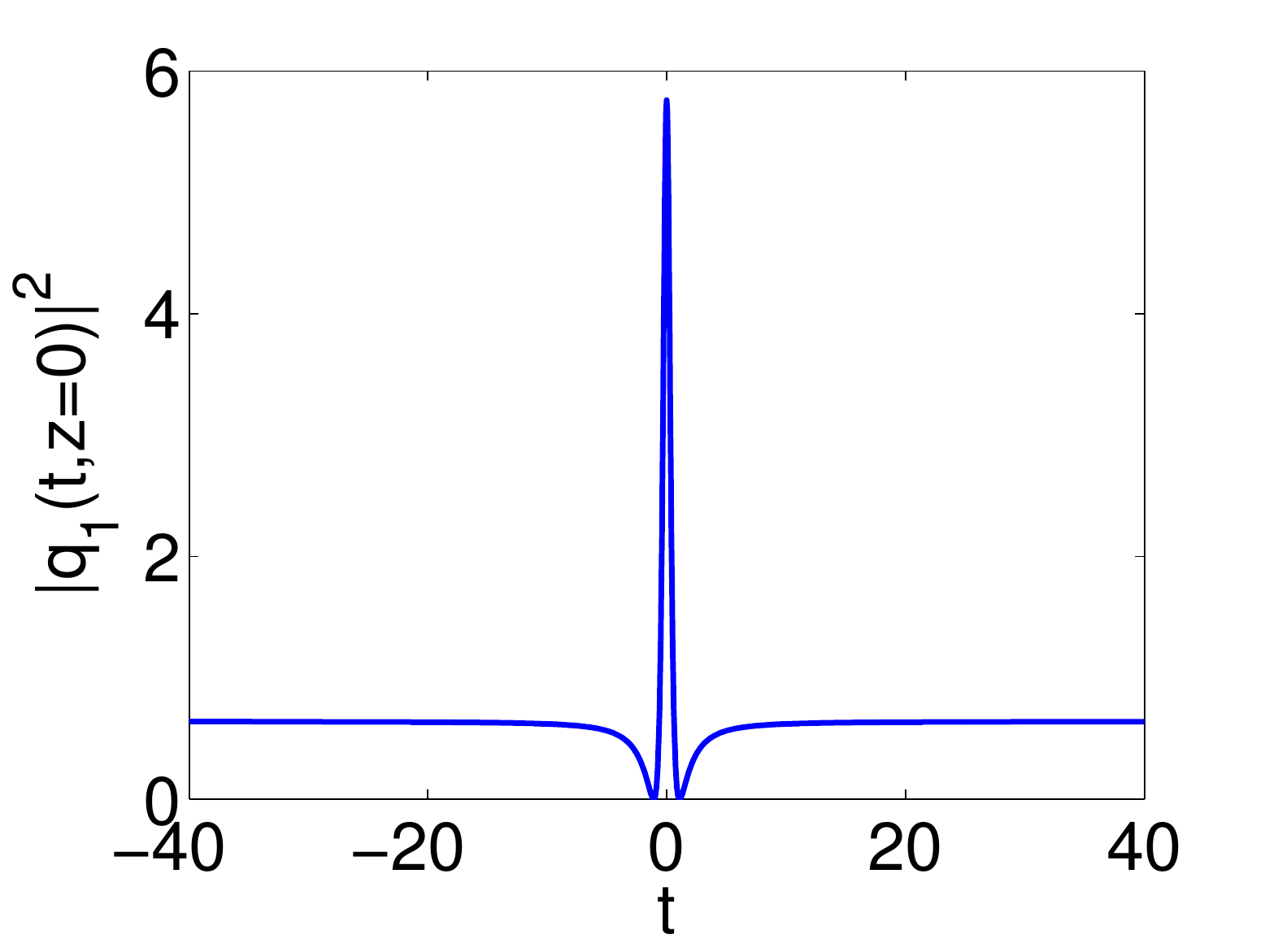}
\label{fig11}
}
\subfigure[][]{\hspace{-0.2cm}
\includegraphics[height=.21\textheight, angle =0]{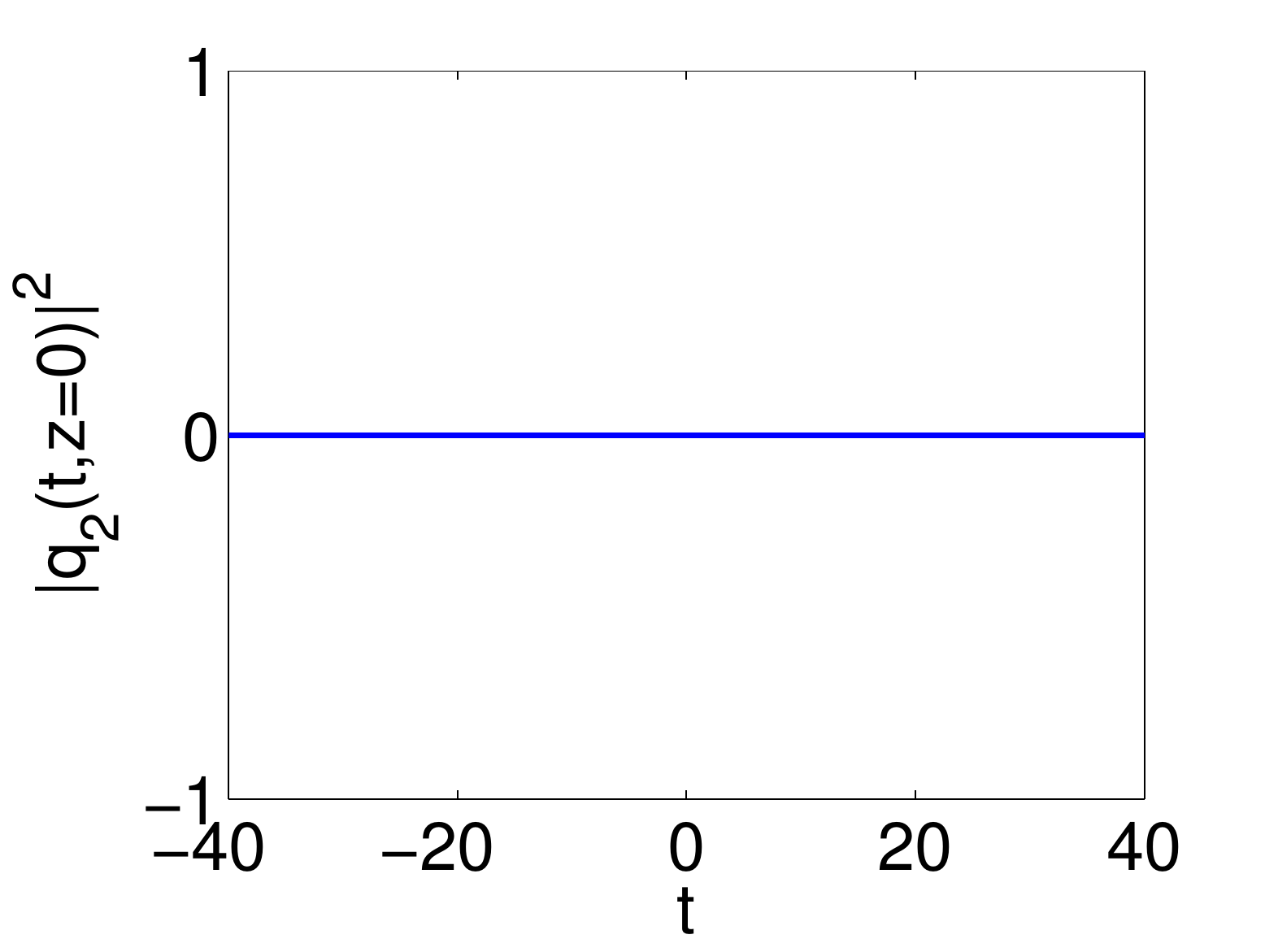}
\label{fig12}
}
}
\mbox{\hspace{-0.2cm}
\subfigure[][]{\hspace{-1.0cm}
\includegraphics[height=.21\textheight, angle =0]{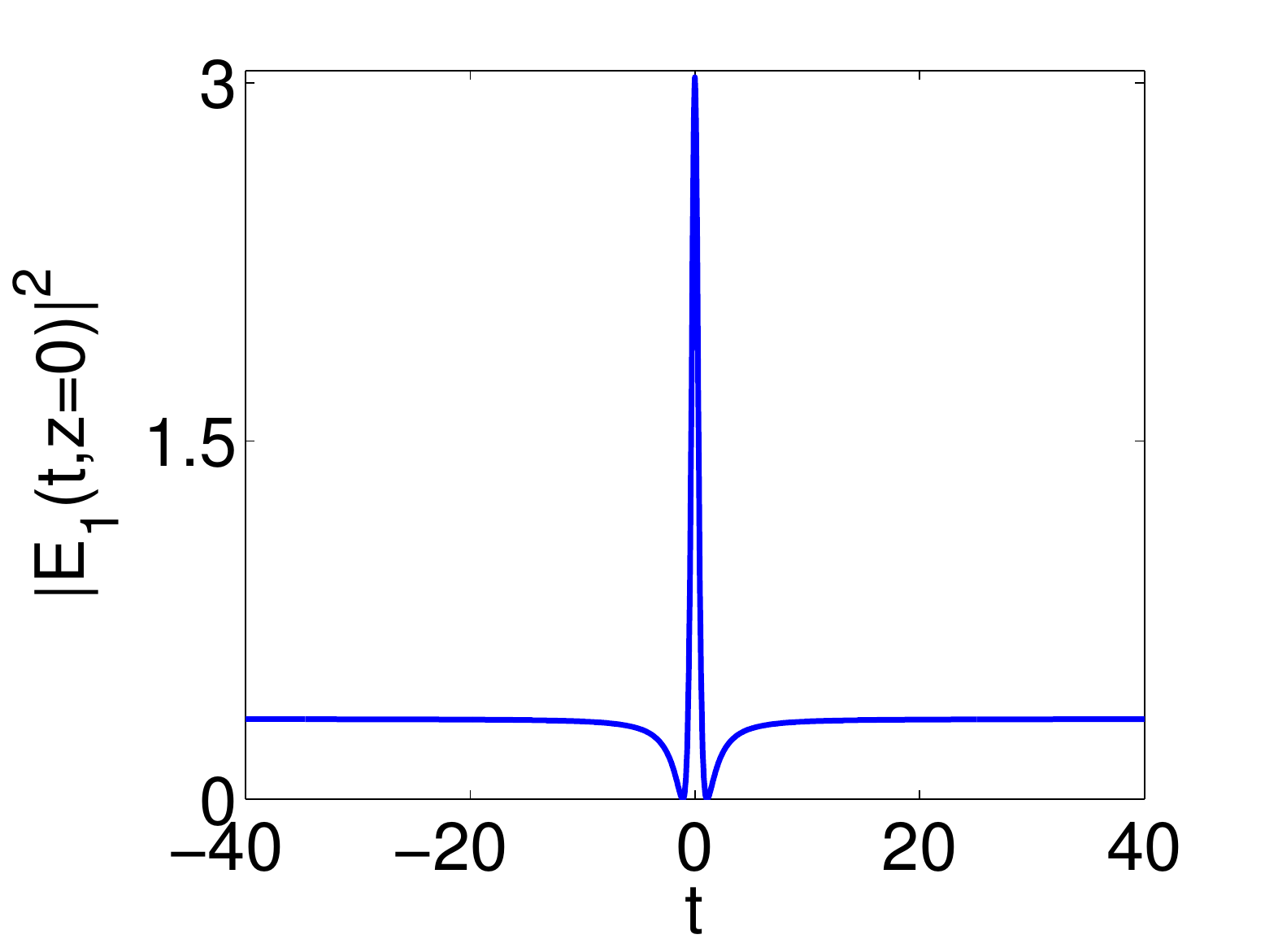}
\label{fig13}
}
\subfigure[][]{\hspace{-0.2cm}
\includegraphics[height=.21\textheight, angle =0]{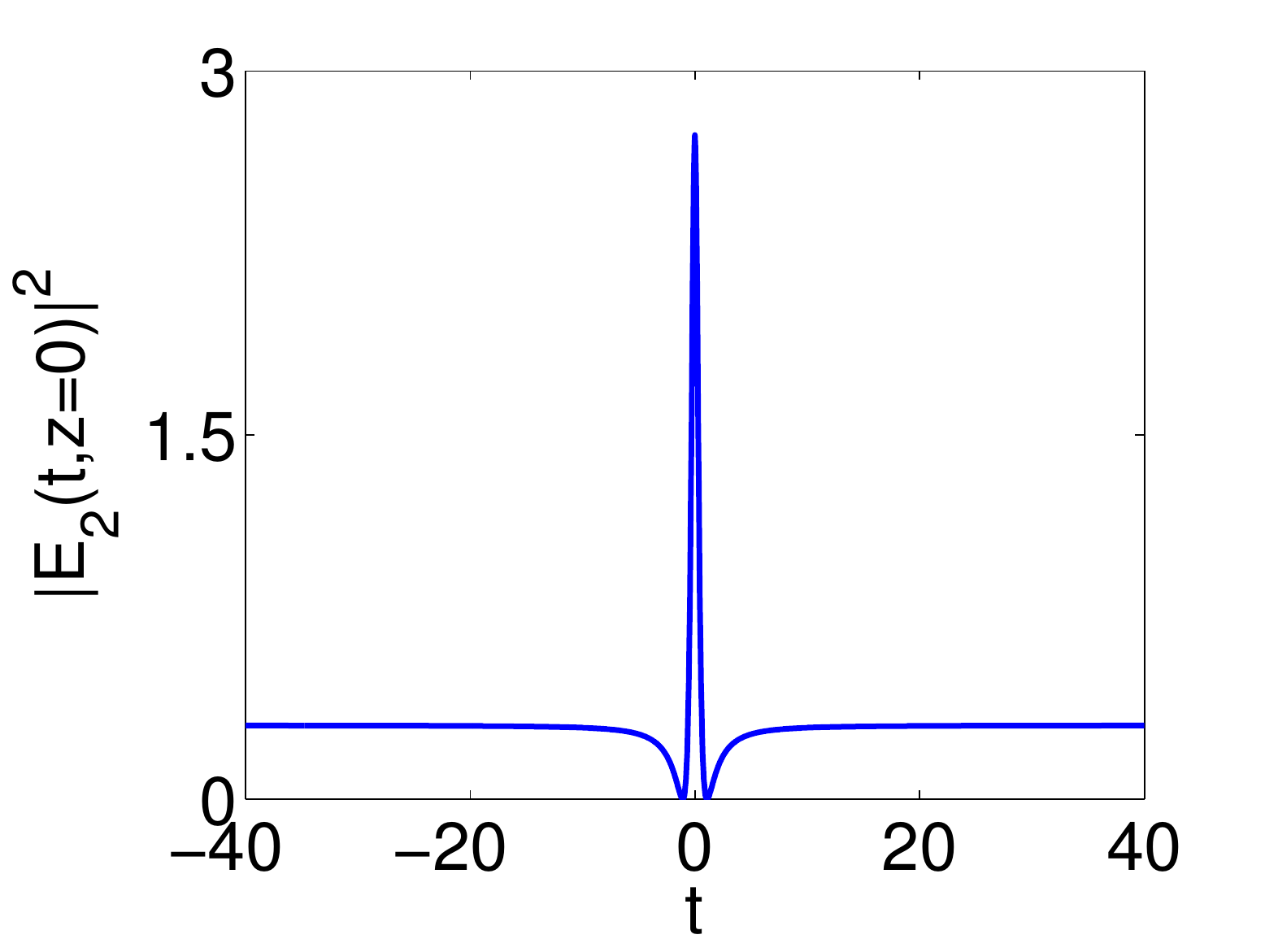}
\label{fig14}
}
}
\mbox{\hspace{-0.2cm}
\subfigure[][]{\hspace{-1.0cm}
\includegraphics[height=.21\textheight, angle =0]{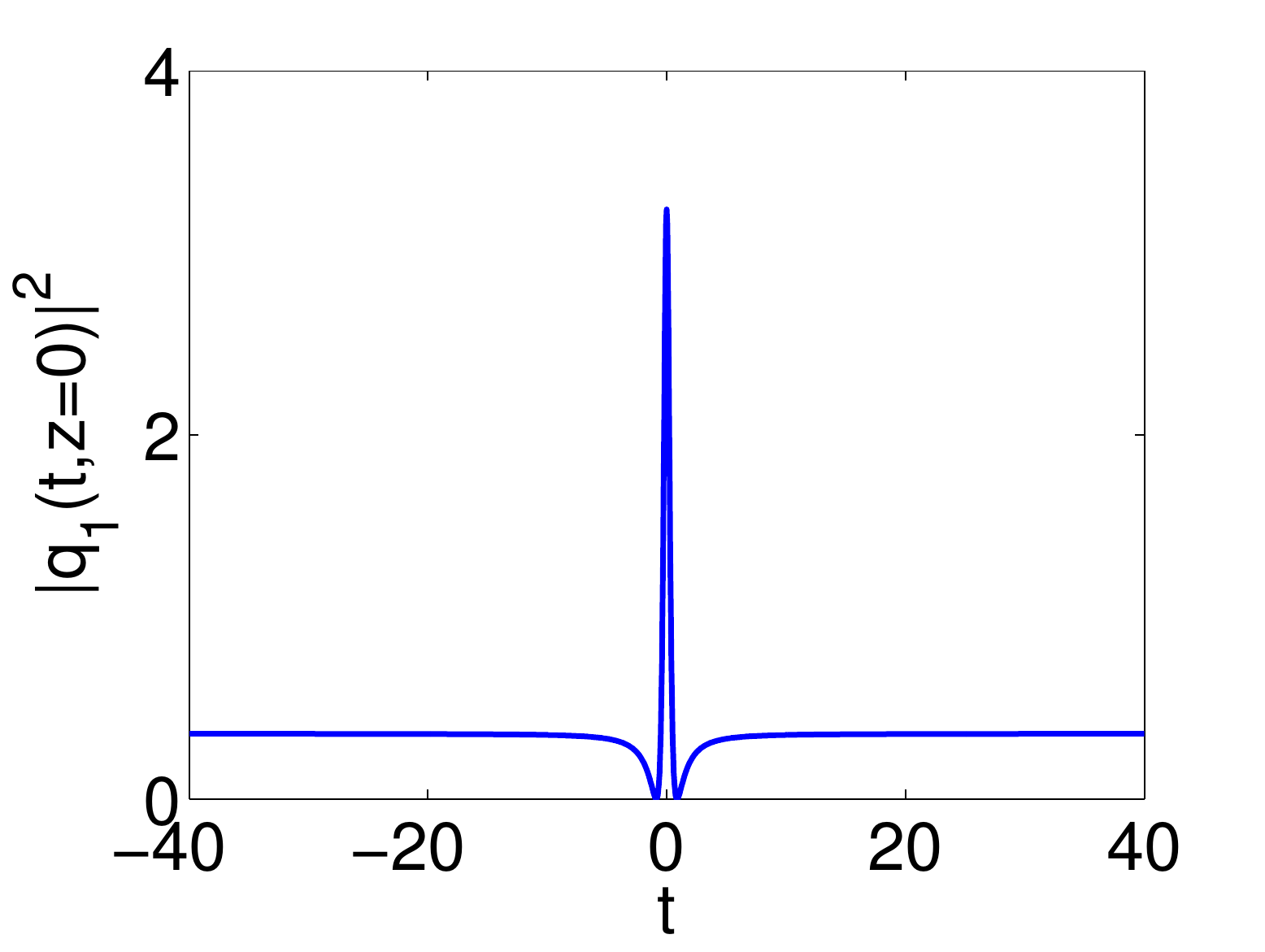}
\label{fig21}
}
\subfigure[][]{\hspace{-0.2cm}
\includegraphics[height=.21\textheight, angle =0]{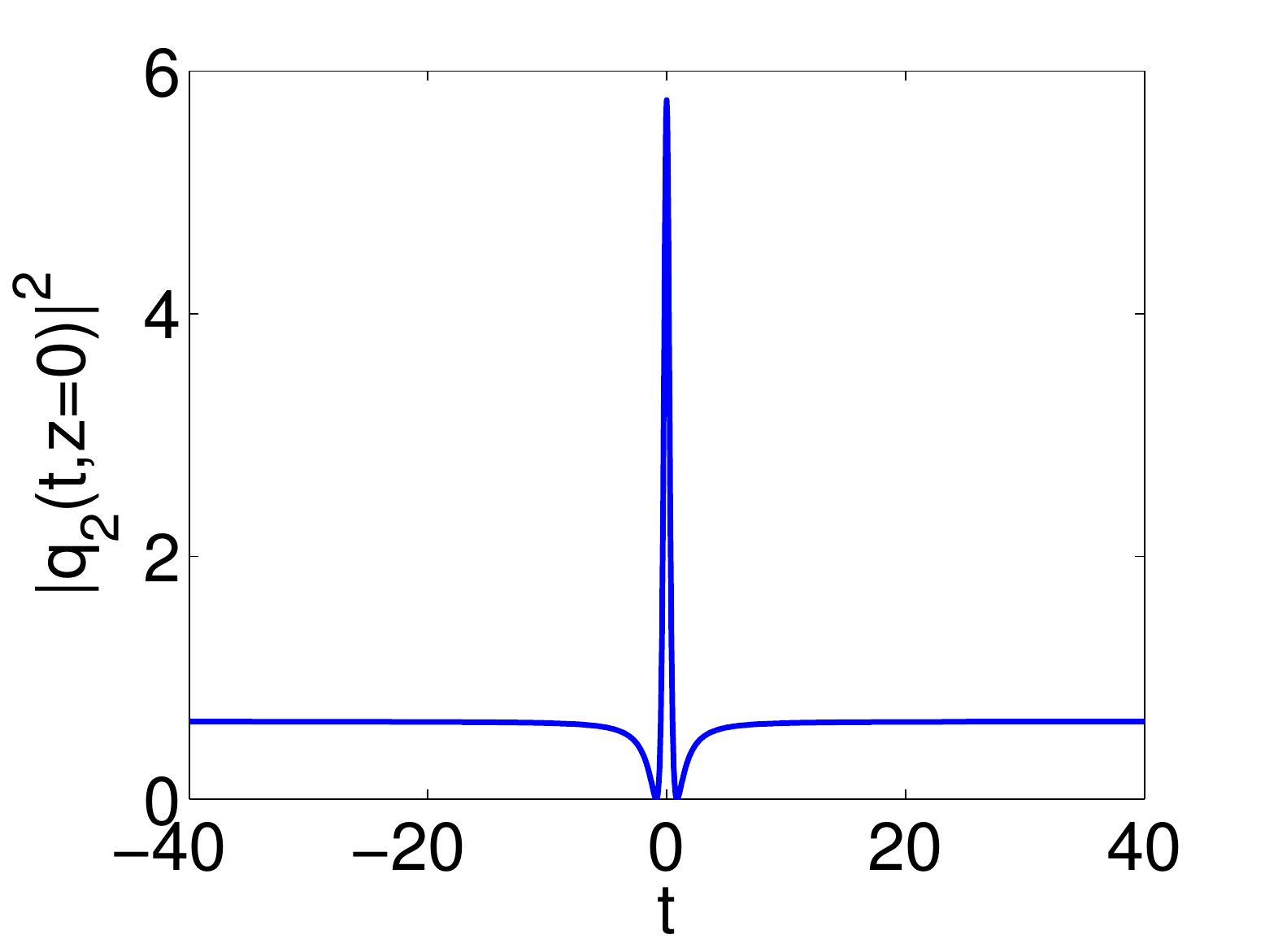}
\label{fig22}
}
}
\mbox{\hspace{-0.2cm}
\subfigure[][]{\hspace{-1.0cm}
\includegraphics[height=.21\textheight, angle =0]{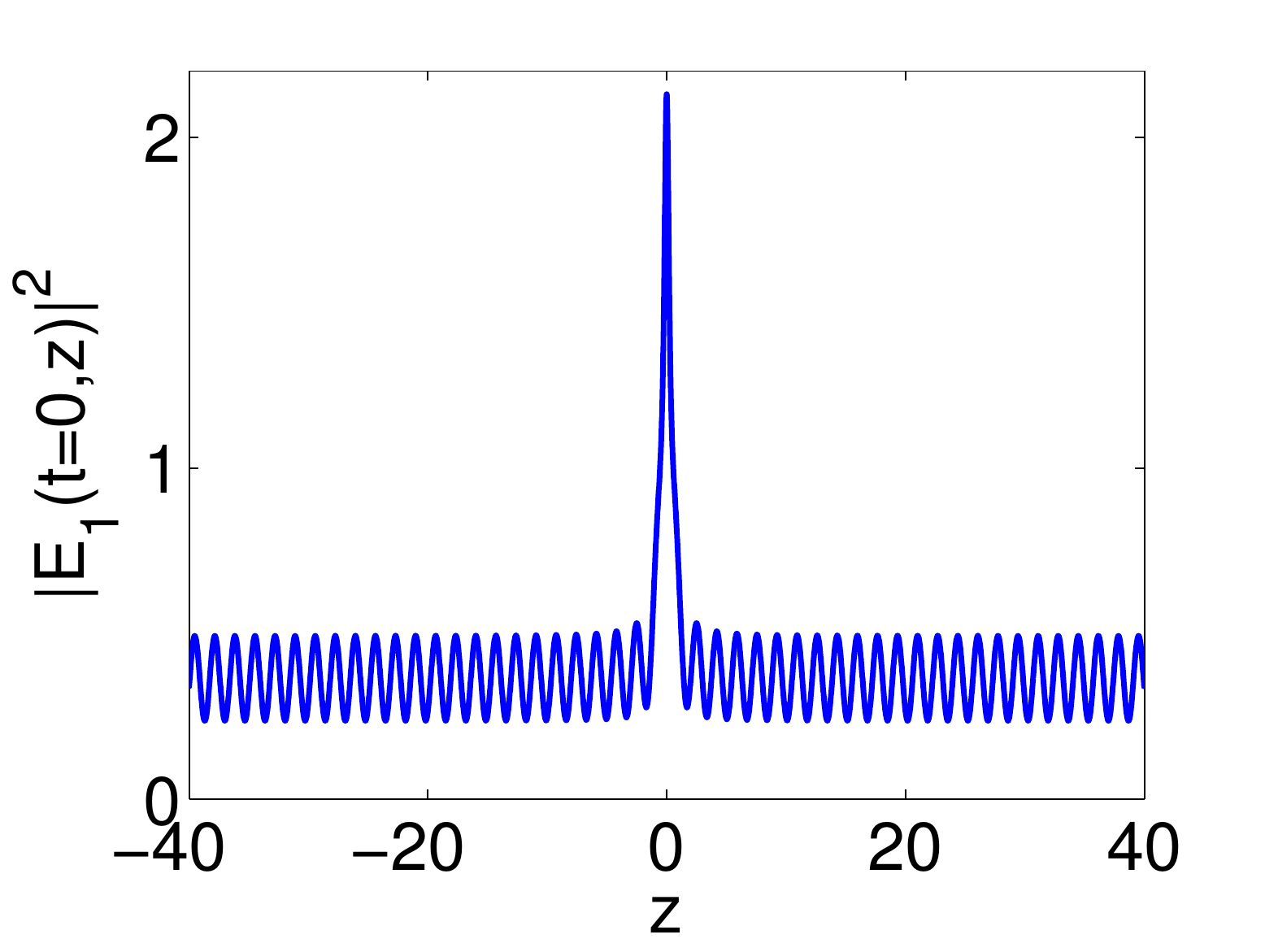}
\label{fig23}
}
\subfigure[][]{\hspace{-0.2cm}
\includegraphics[height=.21\textheight, angle =0]{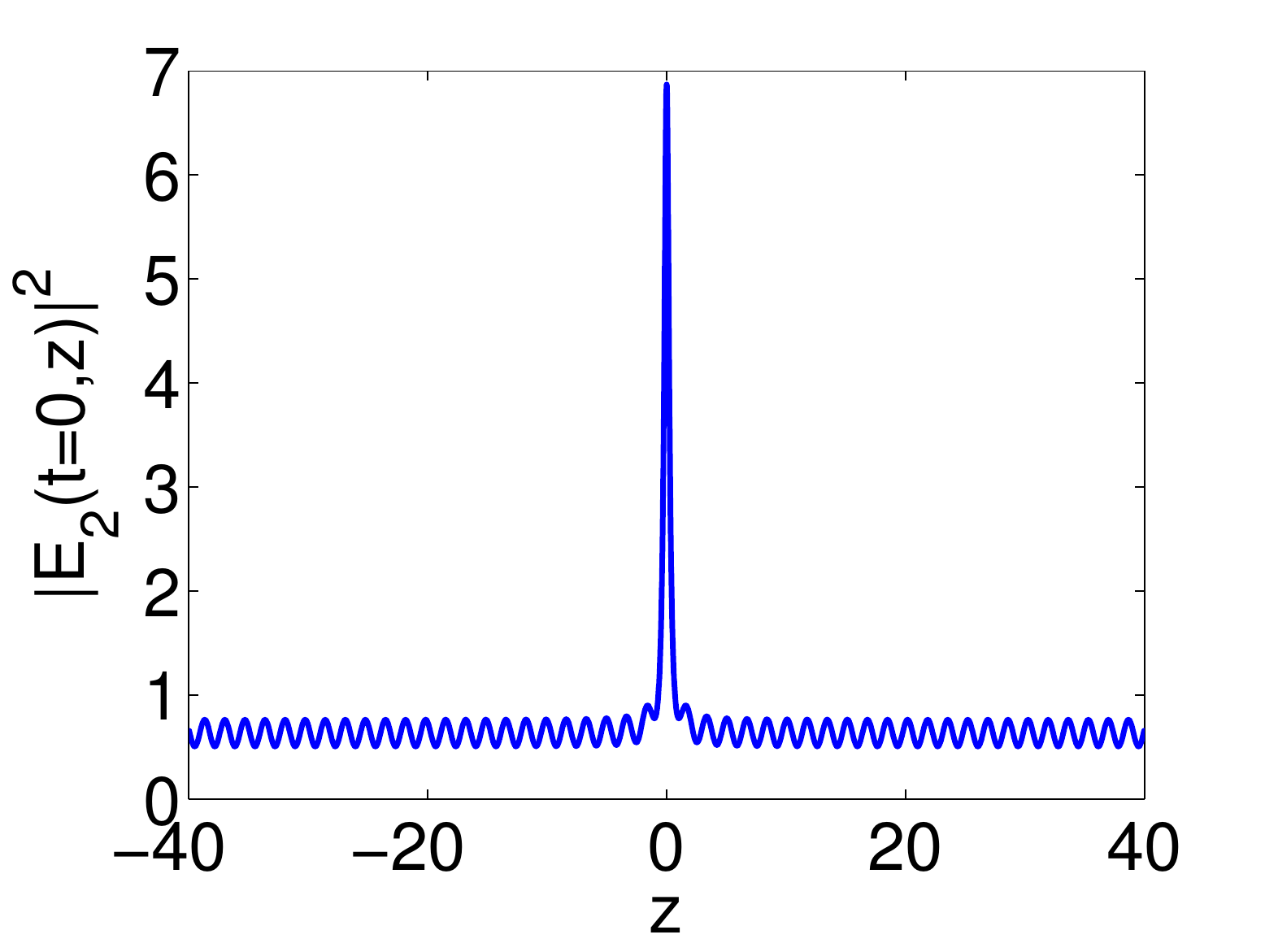}
\label{fig24}
}
}
\end{center}
\caption{(Color online) Exact rogue waves solutions in the absence (first and
third rows) and presence (second and fourth rows) of self- and cross- coupling
parameters. Panels (a) and (b) and (c) and (d) correspond to values of the parameters of 
$a_{1}=0.8$, $a_{2}=0$, $f=0$, $\sigma=1$, and $\alpha=0.05$, whereas 
(e) and (f) and (g) and (h) correspond to values of the parameters of $a_{1}=0.6$, $a_{2}=0.8$,
$f=0$, $\sigma = 0.5$, and $\alpha =1.8$.}
\label{fig11_all}
\end{figure}

\begin{figure}[!pht]
\begin{center}
\vspace{-0.4cm}
\mbox{\hspace{-0.2cm}
\subfigure[][]{\hspace{-0.2cm}
\includegraphics[height=.23\textheight, angle =0]{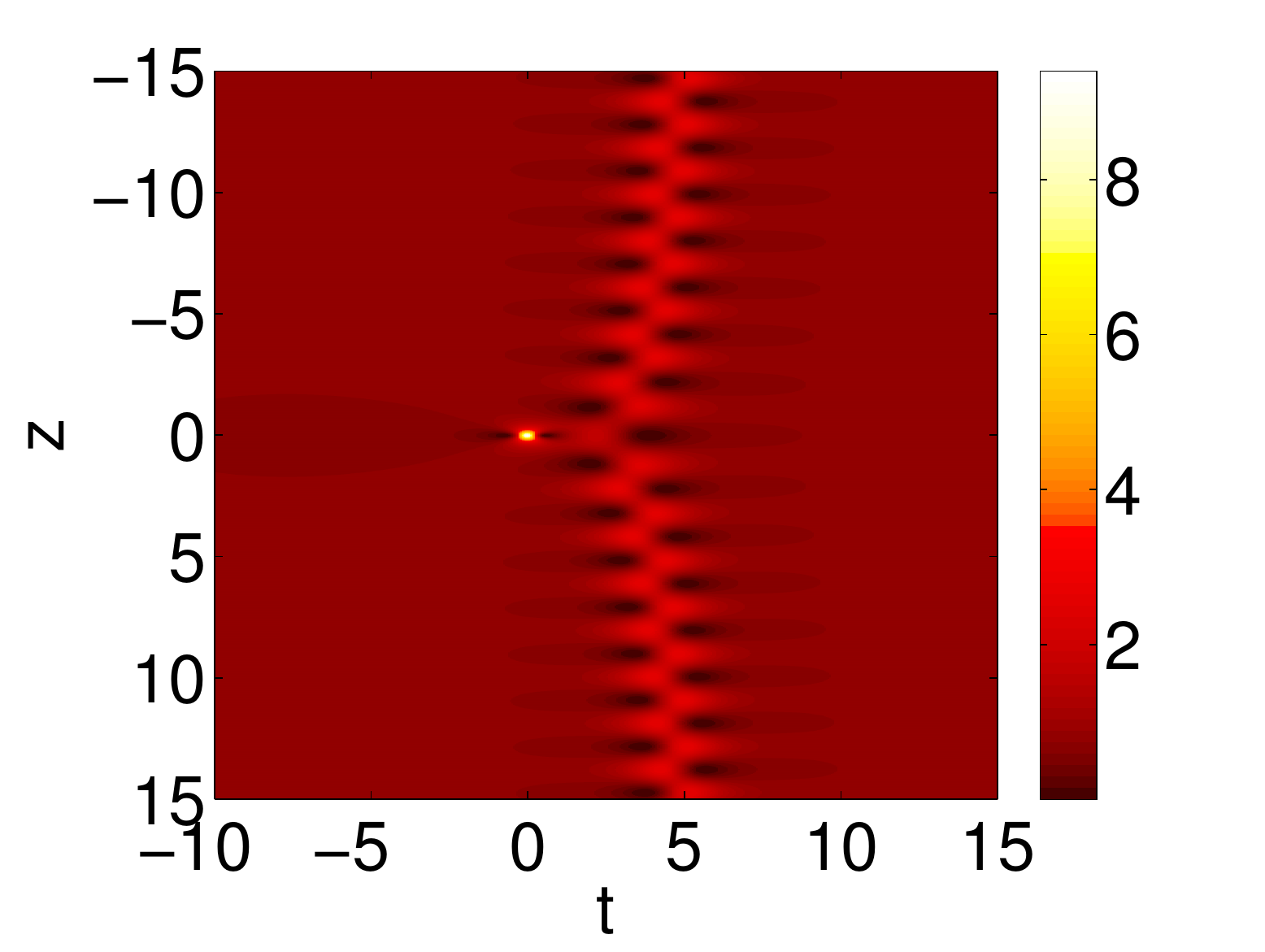}
\label{boom1}
}
\subfigure[][]{\hspace{-0.2cm}
\includegraphics[height=.23\textheight, angle =0]{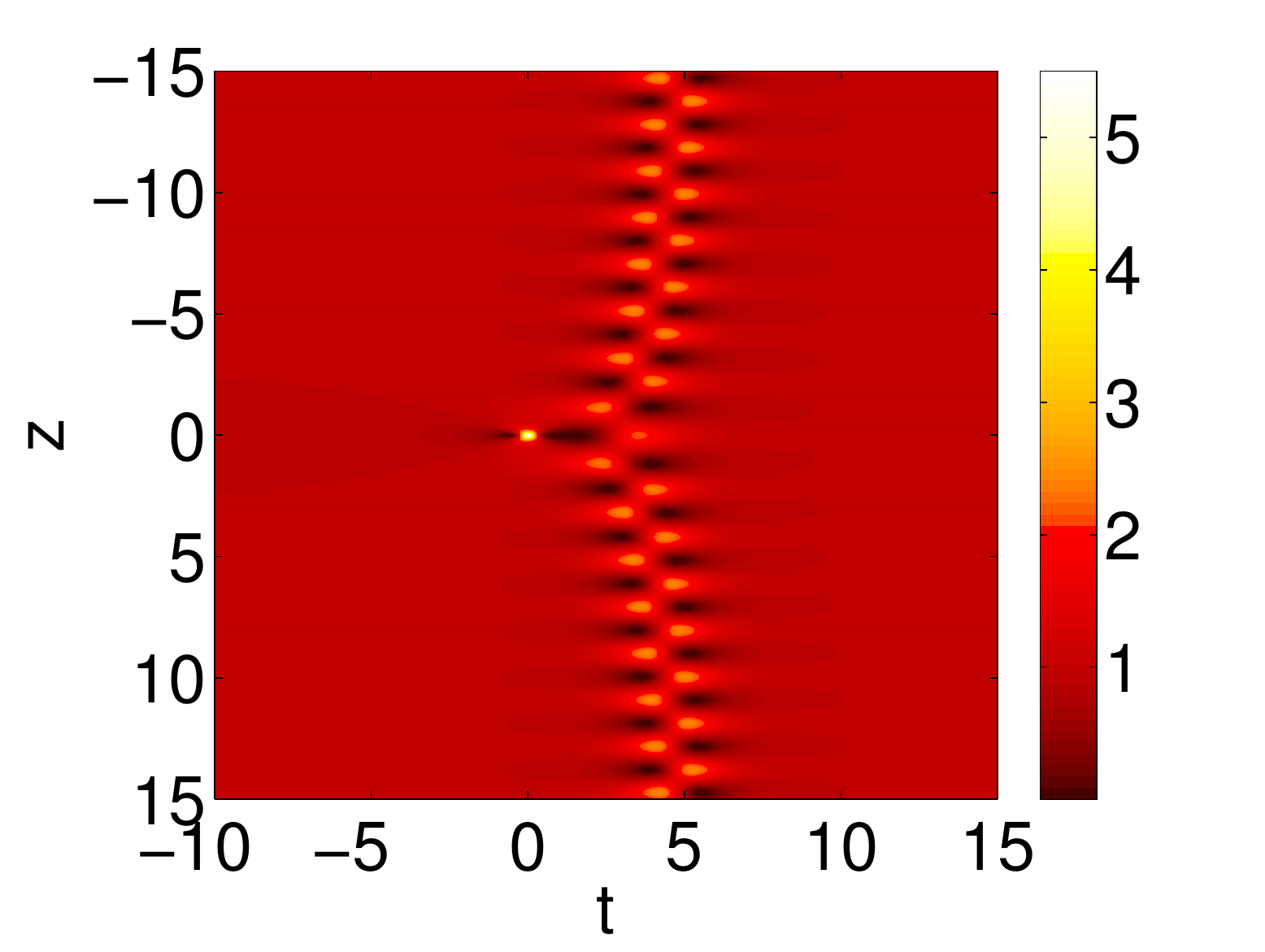}
\label{boom2}
}
}
\end{center}
\caption{(Color online) Contour plots of the density profiles $|E_{1}(t,z)|^{2}$
(left panel) and $|E_{2}(t,z)|^{2}$ (right panel) over space-time corresponding 
to the exact solutions in the form of boomeronic DB solitons in the presence of
self- and cross-coupling parameters with $a_{1}=0$, $a_{2}=1.3$, $f=0.15$, $\sigma = 0.8$,
and $\alpha =0.1$.}
\label{figboom}
\end{figure}

It is interesting to note that when $a_2=0$ and $f=0$, the rogue wave appears in the $q_1$ 
component, whereas the $q_2$ component of the rogue wave vanishes. But in the presence of self- and cross-coupling
parameters even for vanishing $a_2$ and $f$, the rogue waves still appear in both the components $E_1$ 
and $E_2$ without oscillation of the background [see Figs.~\ref{fig11}-\ref{fig14}]. 
On the contrary and for non-vanishing values of $a_1$ and $a_2$ with $f=0$, rogue waves appear 
over an oscillating background [see Figs.~\ref{fig21}-\ref{fig24}]. In fact, it can 
be shown analytically that the oscillations are due to the term: $2a_1a_2\sin\theta \cos\theta e^{2\Gamma z}$ 
which vanishes when $a_1$ (or $a_2$) becomes zero. Furthermore, the linear coupling parameter 
can be advantageously used for controlling the formation of rogue waves. On the other hand, the existence of
the boomeronic type dark-bright (DB) solitons together with rogue waves in the Manakov system requires
non-zero values of $a_1$, $a_2$, and $f$~\cite{degasPRL}. Remarkably, the system (\ref{twogp}) supports 
a slowly moving boomeronic DB soliton co-existing with rogue waves even for $a_1=0$ and $a_2, f\neq0$. 
The latter scenario is depicted in Fig.~\ref{figboom}. It should be noted that as $a_1$, $a_2$, and $f$
vary, we observe that if the amplitude of the DB soliton is increased  (by tuning appropriately these 
parameters), then the amplitude of the rogue wave will be decreased and vice versa. Having discussed 
the constant (coupling) coefficient case, we now proceed with the case of modulated (in the evolution 
variable) system parameters.

\subsection{Nonlinearity management of rogue waves}

We now turn our focus to an inhomogeneous generalization of the NLS system of
Eqs.~(\ref{twogp}) with variable nonlinearity coefficient; this model
is expressed in dimensionless form as follows:
\bea
i E_{1,z}+E_{1,tt}+2\gamma(z)\sum_{l=1}^2|E_l|^2E_1+\left[\upsilon(t,z)+\alpha\right]E_1+\sigma E_2=0,\nonumber\\
i E_{2,z}+E_{2,tt}+2\gamma(z)\sum_{l=1}^2|E_l|^2E_2-\left[\alpha-\upsilon(t,z)\right]E_2+\sigma E_1=0.
\label{inhomogeneous}
\eea
Our motivation, as indicated also above, stems from the possibility
to modulate $\gamma$ in the evolution variable by means of either
Feshbach resonances in BEC~\cite{mFRM,oFRM} or by
means of layered optical media in nonlinear optics~\cite{psaltis}. Additionally,
Eq.~(\ref{inhomogeneous}) features an external potential which is given by $\upsilon(t,z)$. 
This potential will be assumed to be parabolic in the transverse variable $t$ 
(accounting for a linear parabolic refractive index profile in optics, or a harmonic 
trapping potential in BECs), while it will also be assumed to be modulated in 
the evolution variable $z$, i.e., $\upsilon(t,z)=F(z) t^2/2$. 
This way, the parabolic refractive index (in optics) or the parabolic trap (in BECs) is
also modulated in the propagation direction (in optics) or in time (in BECs).
Examples of such modulations have again arisen in recent experimental
efforts in optics~\cite{szameit}, while they have been widely used in the physics of BECs \cite{emergent}.

Now, we introduce the following rotational transformation to Eq.~(\ref{inhomogeneous}):
\[
\begin{pmatrix}
E_{1} \\ E_{2}
\end{pmatrix}
=\begin{pmatrix}
\cos{\theta} & - \sin{\theta} \\
\sin{\theta} &   \cos{\theta}
\end{pmatrix}
\begin{pmatrix}
 \phi_{1}e^{i\Gamma z} \\
 \phi_{2}e^{-i\Gamma z}
 \end{pmatrix},
 \label{rabi1}
\]
where $\Gamma$ = $\sqrt{\sigma^2+\alpha^2}$ and $\theta = \frac{1}{2}\tan^{-1}(\sigma/\alpha)$. The resulting equation
can be rewritten in the following manner:
\bea
i\phi_{j,z}+\phi_{j,tt}+2\gamma(z)\sum_{\ell=1}^2 |\phi_\ell|^2\phi_j+\upsilon(t,z)\phi_j=0,\quad j=1,2.
\label{phiequation}
\eea
We introduce the similarity transformation
\bes\bea
\phi_{j}(t,z)=\zeta_1\sqrt{\gamma(z)}~e^{i\varphi(t,z)}q_j[T(t,z),Z(z)],\quad j=1,2,
\label{ansatz}
\eea
together with a quadratic ansatz for the phase:
\bea
\varphi = -\frac{1}{4}\left[\frac{d}{dz}(\ln \gamma)\right]t^2 + \zeta_2 \zeta_1^2\left(\gamma t -  \zeta_2 \zeta_1^2\int \gamma^2 dz\right),
\label{phase}
\eea
and the coordinate transformation
\bea
&&T= \zeta_1 \left(\gamma t - 2\zeta_2 \zeta_1^2\int \gamma^2 dz\right),\label{xzcoordinate1}\\
&&Z= \zeta_1^2 \int \gamma^2 dz,
\label{xzcoordinate2}
\eea\ees
to obtain a standard self-focusing integrable Manakov system:
\bes\bea
iq_{j,Z}+q_{j,TT}+2\sum_{\ell=1}^2|q_\ell|^2q_j=0,\quad j=1,2,
\eea
along with the following integrability condition:
\bea
F(z)=-\frac{1}{2\gamma}\frac{d^2\gamma}{dz^2}+\frac{1}{\gamma^2}\left(\frac{d\gamma}{dz}\right)^2.
\eea\ees
Now, we can reconstruct the exact form of the rogue wave solution of Eq.~(\ref{inhomogeneous}):
\bes\bea
E_1(t,z) = \rho\left(\cos\theta e^{i(\Gamma z+\varphi)}(\bar{G}a_1+\bar{M}a_2)- \sin\theta e^{i(-\Gamma z+\varphi)}(\bar{G}a_2-\bar{M}a_1)\right), \label{rw_E1}\\
E_2(t,z) = \rho\left(\sin\theta e^{i(\Gamma z+\varphi)}(\bar{G}a_1+\bar{M}a_2)+ \cos\theta e^{i(-\Gamma z+\varphi)}(\bar{G}a_2-\bar{M}a_1)\right), \label{rw_E2}
\eea\ees
where $\rho = \frac{\zeta_1\sqrt{\gamma}}{\bar{F}} e^{2i\omega Z}$, $\bar{G} = (3/2) - 8\omega^2Z^2-2a^2T^2+8i\omega Z+|f|^2e^{2aT}$, 
$\bar{M} = 4f(aT-2i\omega Z-\frac{1}{2})e^{aT+i\omega Z}$, $\bar{F} = (1/2)+8\omega^2Z^2+2a^2T^2+|f|^2e^{2aT}$ 
and the coordinates $T$ and $Z$ are given by Eqs.~(\ref{xzcoordinate1})-(\ref{xzcoordinate2}).


\subsubsection{\it Periodically modulated nonlinearity coefficient}
For purposes of illustration, we now give some specific, yet realistic
examples of the nonlinear coefficient and linear potential prefactor variations.
First we choose a periodic form of the nonlinearity coefficient, namely,
\bes\bea
\gamma(z) = 1+ \varepsilon\cos(z),
\label{non-coefficient}
\eea
where $\varepsilon$ is a real arbitrary parameter. Periodic variations
of the nonlinearity coefficient are commonly considered both in BECs~\cite{BEC,pelinov} and in layered
optical media~\cite{towers,psaltis}.
The accompanying evolutionary modulation of the trap frequency reads:
\bea
F(z) = \frac{\varepsilon[\cos(z)+\varepsilon\cos^2(z)+2\varepsilon \sin^2(z)]}{2[1+\varepsilon \cos(z)]^2}.
\label{trap}
\eea\ees

\begin{figure}[!pht]
\begin{center}
\vspace{-0.1cm}
\mbox{\hspace{-0.1cm}
\subfigure[][]{\hspace{-0.2cm}
\includegraphics[height=.21\textheight, angle = 0]{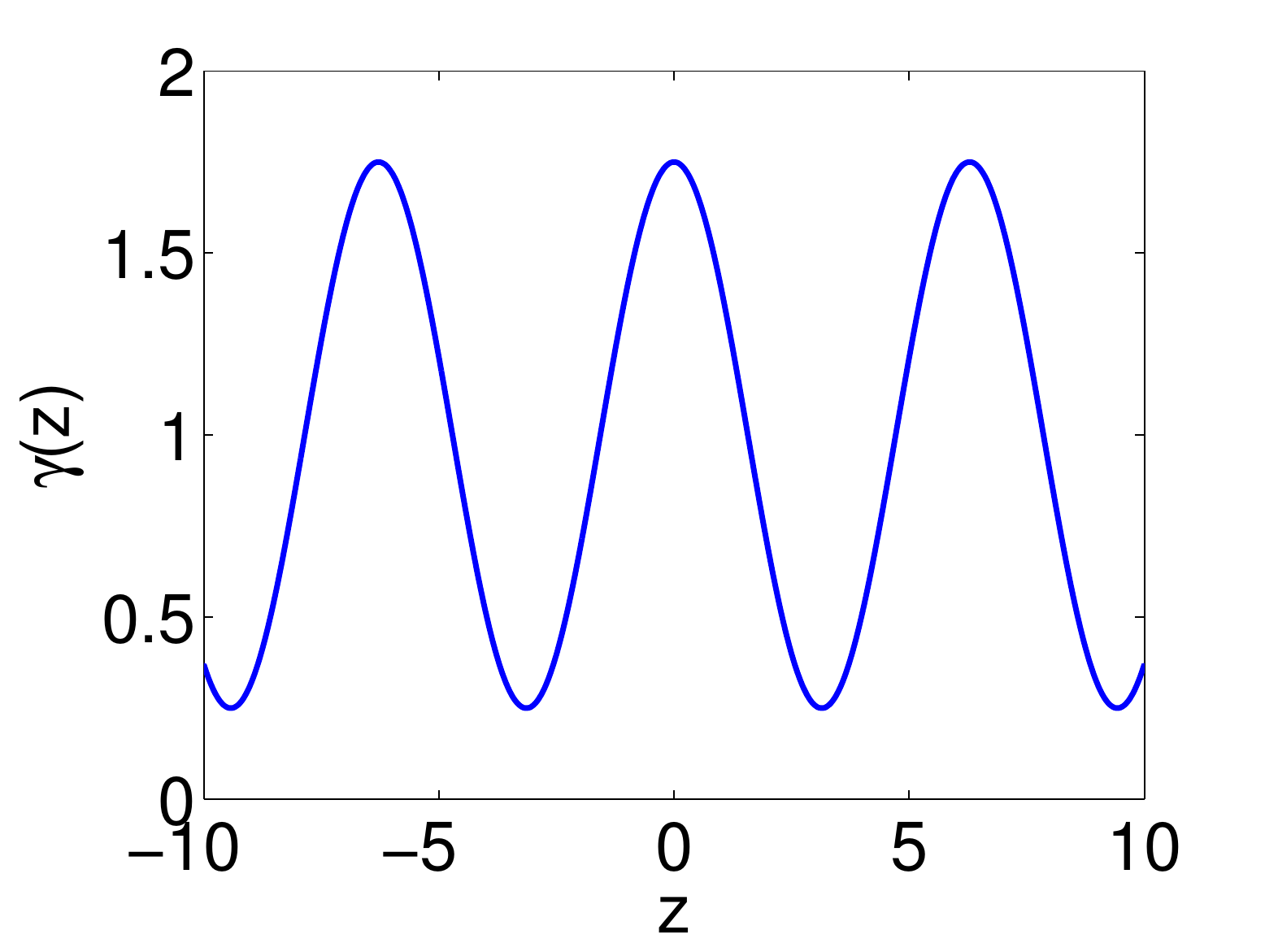}
\label{fig1a}
}
\subfigure[][]{\hspace{-0.2cm}
\includegraphics[height=.21\textheight, angle = 0]{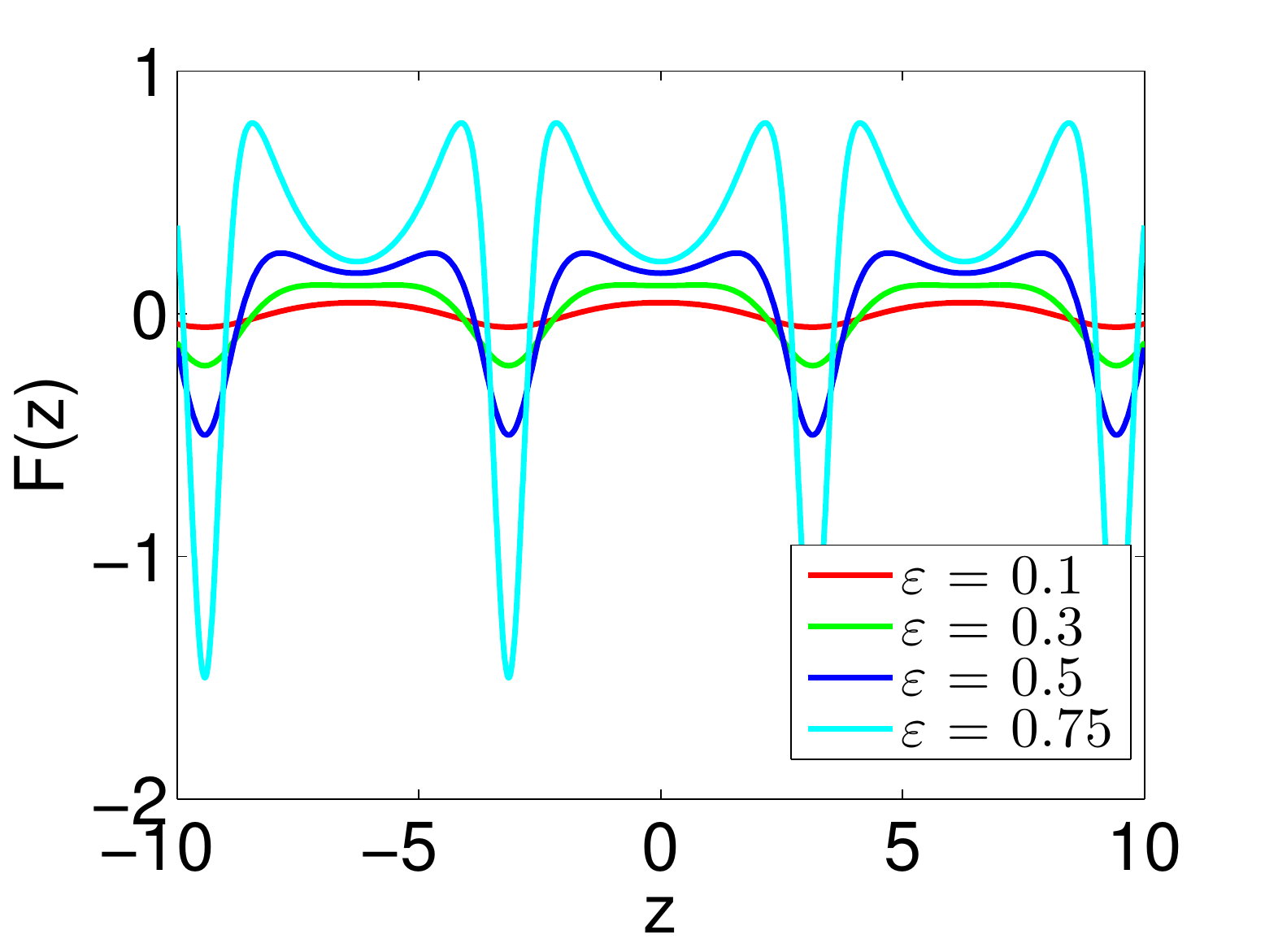}
\label{fig1b}
}
}
\end{center}
\vspace{-0.7cm}
\caption{(Color online) Typical form of variable nonlinearity coefficient
$\gamma(z)$ for $\varepsilon=0.75$ and trap frequency $F(z)$ for various 
values of $\varepsilon$.}
\label{fig1}
\end{figure}

Figure~\ref{fig1} depicts the profiles of the variable nonlinearity coefficient $\gamma(z)$
and trap frequency $F(z)$ given by Eqs.~(\ref{non-coefficient}) and (\ref{trap}), respectively.
In this case, in the absence of cross- and self- coupling parameters $\sigma$ and $\alpha$,
the Peregrine bump is not deformed from the original structure, while the 
dark-bright soliton's central position is varied with respect to the propagation
coordinate $z$, as well as the amplitude increases in both
$\phi_1$ and $\phi_2$ components. In the presence of coupling parameters, the
dark-bright
solitons may exhibit oscillations (creeping soliton), e.g., for a
smaller value of $\sigma$ and larger value of $\alpha$, whereas for larger values
of $\sigma$ and smaller $\alpha$'s, the boomeronic behavior is encountered.

\subsubsection{\it Kink-like nonlinearity coefficient}
We also chose to examine another form of variable nonlinearity coefficient, 
 enabling the transition between two distinct (constant) values
 of $\gamma$, namely,
\bes\bea
\gamma(z) = 2+\tanh(\varepsilon z),
\label{eq12a}
\eea
where $\varepsilon$ is, again, a real arbitrary parameter, while the associated form of trap frequency is:
\bea
F(z) = \frac{\varepsilon^2\mbox{sech}^2(\varepsilon z)[1+2~\tanh(\varepsilon z)]}{[2+\tanh(\varepsilon z)]^2}.
\label{eq12b}
\eea\ees
In this case, we envision a nonlinearity that is rapidly varied from
one value to another as is often done, e.g., in atomic condensates
to explore the response of the condensate to such variations~\cite{hulet}.

\begin{figure}[!pht]
\begin{center}
\vspace{-0.1cm}
\mbox{\hspace{-0.1cm}
\subfigure[][]{\hspace{-0.2cm}
\includegraphics[height=.21\textheight, angle = 0]{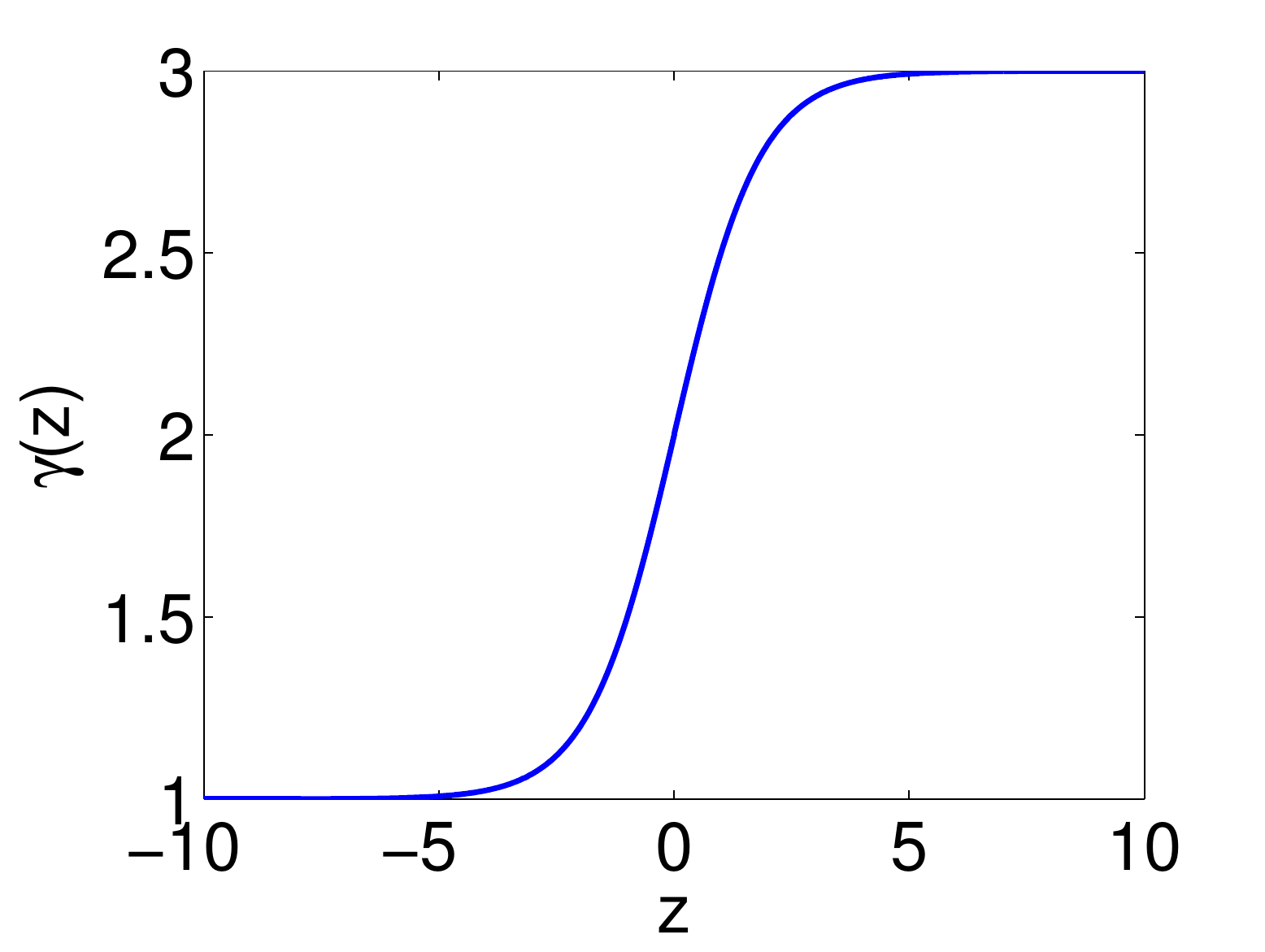}
\label{fig2a}
}
\subfigure[][]{\hspace{-0.2cm}
\includegraphics[height=.21\textheight, angle = 0]{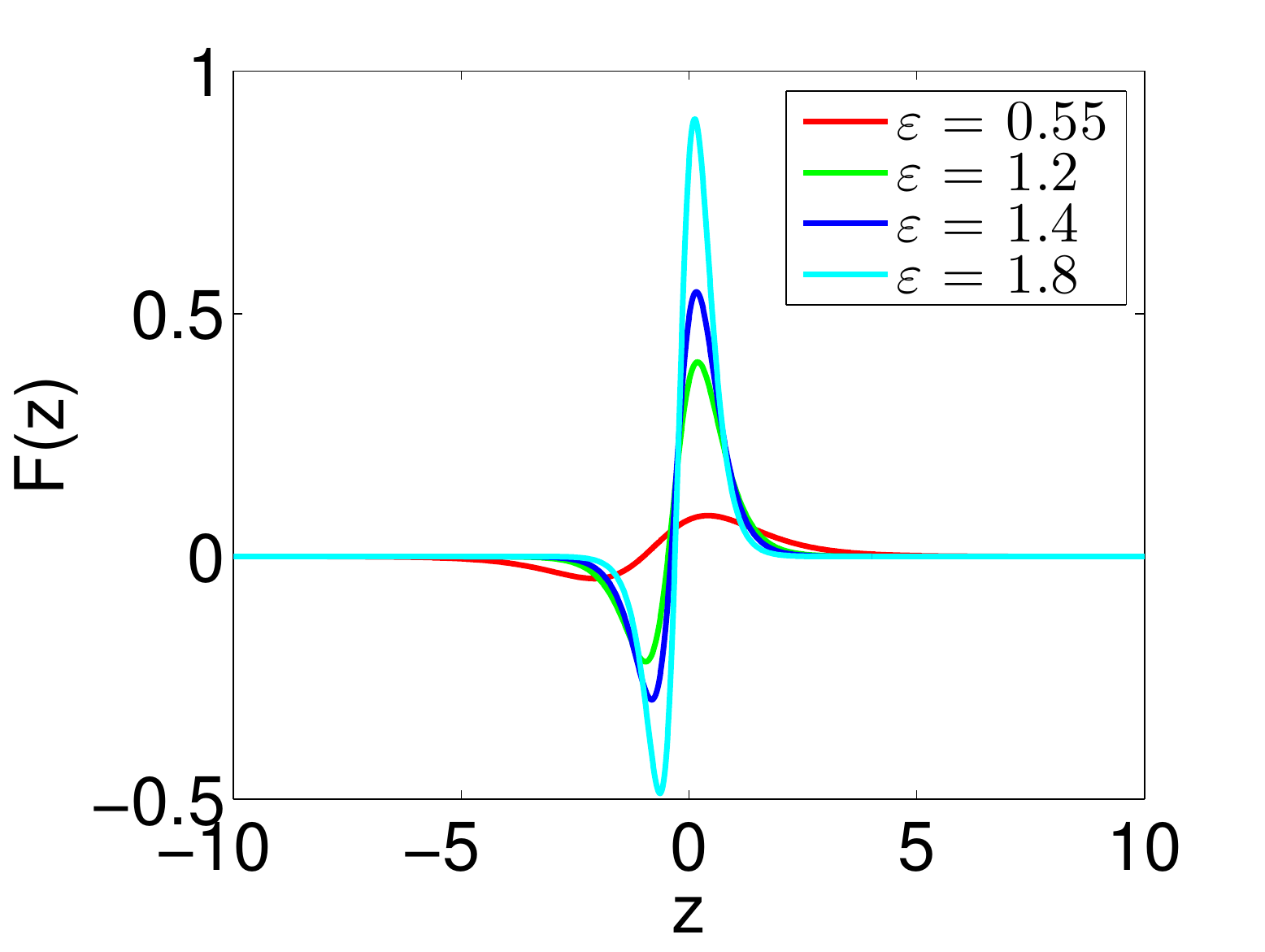}
\label{fig2b}
}
}
\end{center}
\vspace{-0.7cm}
\caption{(Color online) Profiles of nonlinearity $\gamma(z)$ (for $\varepsilon=0.55$)
and trap frequency $F(z)$ given by Eqs.~(\ref{eq12a}) and (\ref{eq12b}), respectively.}
\label{fig2}
\end{figure}

The graphical form of the nonlinearity coefficient and of the trap frequency
as a function of the evolution variable are shown in Fig.~\ref{fig2}.
The modulation of the confining potential $F(z)$ is an asymmetric
localized pulse of finite duration;  see the right panel of the figure.

In observing the two components in this case,
we see that the rogue waves are located on a modified
(kink-like, as the kink-like variation in the nonlinearity and linear
potential sets in) background.
Furthermore,
given the functional form of $\gamma(z)$=2+$\tanh(\varepsilon z)$, the soliton
expansion begins
at the initial value of $z$. On the contrary, when $z>0$, the DB
soliton is compressed into a higher amplitude
and narrower one.

Having provided the above analytical descriptions for the two different
(physically relevant)
cases of $\gamma(z)$ [and $F(z)$],
we now turn to a series of numerical results. These
will  help elucidate
the states considered (vector rogue waves and rogue-boomeron solution
complexes) and their dynamical robustness and observability.

\section{Computational Analysis}
\subsection{Numerical method}
In what follows, numerical results are presented for the direct numerical integration
of Eqs.~(\ref{inhomogeneous}) over the evolution variable by
discretizing the $t$ variable, i.e.,
converting the system of partial differential equations (PDEs) into a system of ordinary differential equations (ODEs).
In the same spirit as in Ref.~\cite{He2014577}, we explore the
robustness of the exact solutions describing both Peregrine and interacting
Peregrine-DB solitons of Eqs.~(\ref{inhomogeneous}) [by tuning both the
coupling parameters and
those involved in Eqs.~(\ref{rw_E1}) and (\ref{rw_E2}) appropriately], under
the presence of numerically induced perturbations (caused by round-off and truncation errors). Briefly,
the numerical method can be summarized as follows.

At first, the second-order spatial derivatives with respect to $t$ arising
in Eqs.~(\ref{inhomogeneous})
are replaced by fourth-order central difference formulas on a uniform spatial
grid (with half-width $L$) consisting of $N$ points and lattice spacing $\Delta{t}$ with
$t_{j}=-L+2jL/(N+1)$, respectively.
In all numerical experiments presented herein, the grid's half-width and 
resolution are chosen to be $L=500$ and $\Delta{t}=0.05$ (except for the case of Fig.~\ref{fig3} where $L=300$),
respectively, and both are kept fixed.
It should be noted that although the fourth-order central difference
formulas for the spatial derivatives with respect to $t$ require a wide five-point stencil,
one point away both from the left and right boundaries, we use instead second-order central difference formulas.
As far as boundary conditions are concerned, no-flux ones are applied, namely,
$E_{j,t}(t,z)|_{t=-L}=E_{j,t}(t,z)|_{t=L}=0$, with $j=1,2$. The latter are
coupled into the internal
discretization scheme using second-order forward and backward difference
formulas, respectively.

The temporal integration of the underlying system of ODEs is performed using
the
Dormand and Prince method (DOP853) with an automatic time-step adaptation
procedure (cf. the Appendix
in Ref.~\cite{Hairer}). Furthermore, our numerical results are corroborated using the standard fourth-order
Runge-Kutta method (RK4) with fixed time step (typically $\Delta{z}=10^{-5}$ or $10^{-6})$, although
numerical results are presented using only the DOP853 method throughout this section;
we have nevertheless confirmed that the results of RK4 are
essentially identical. We initialize
the dynamics using the exact vector rogue wave solutions given by Eqs.~(\ref{rw_E1}) and (\ref{rw_E2})
adjusted to the particular cases presented below (see also Table~\ref{cnls_numer_summary} for details about
parameter values), and consider both periodic and kink-like modulated
nonlinearities given by
Eqs.~(\ref{non-coefficient}) and (\ref{eq12a}), respectively. Finally, our numerical integrator
is initialized (in all cases studied) at $z_{i}=-3$, since the rogue waves in the vector system
appear around $z=0$ and this way, we keep track of the dynamical
evolution of the solutions
shortly before high amplitude coherent structures arise.

\subsection{Numerical results}
\begin{figure}[!pht]
\begin{center}
\vspace{-0.7cm}
\mbox{\hspace{-0.2cm}
\subfigure[][]{\hspace{-1.0cm}
\includegraphics[height=.21\textheight, angle =0]{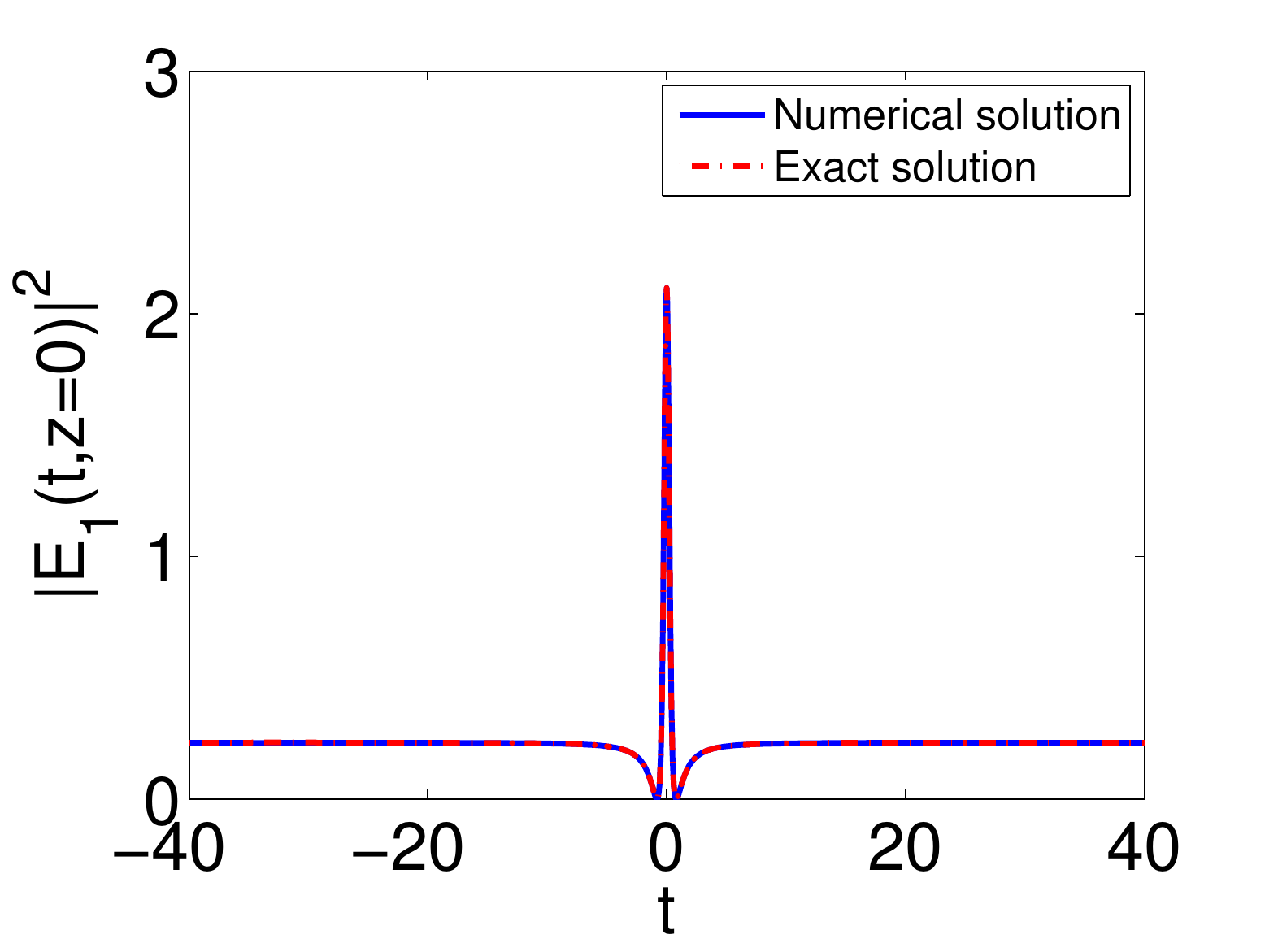}
\label{fig3a}
}
\subfigure[][]{\hspace{-0.2cm}
\includegraphics[height=.21\textheight, angle =0]{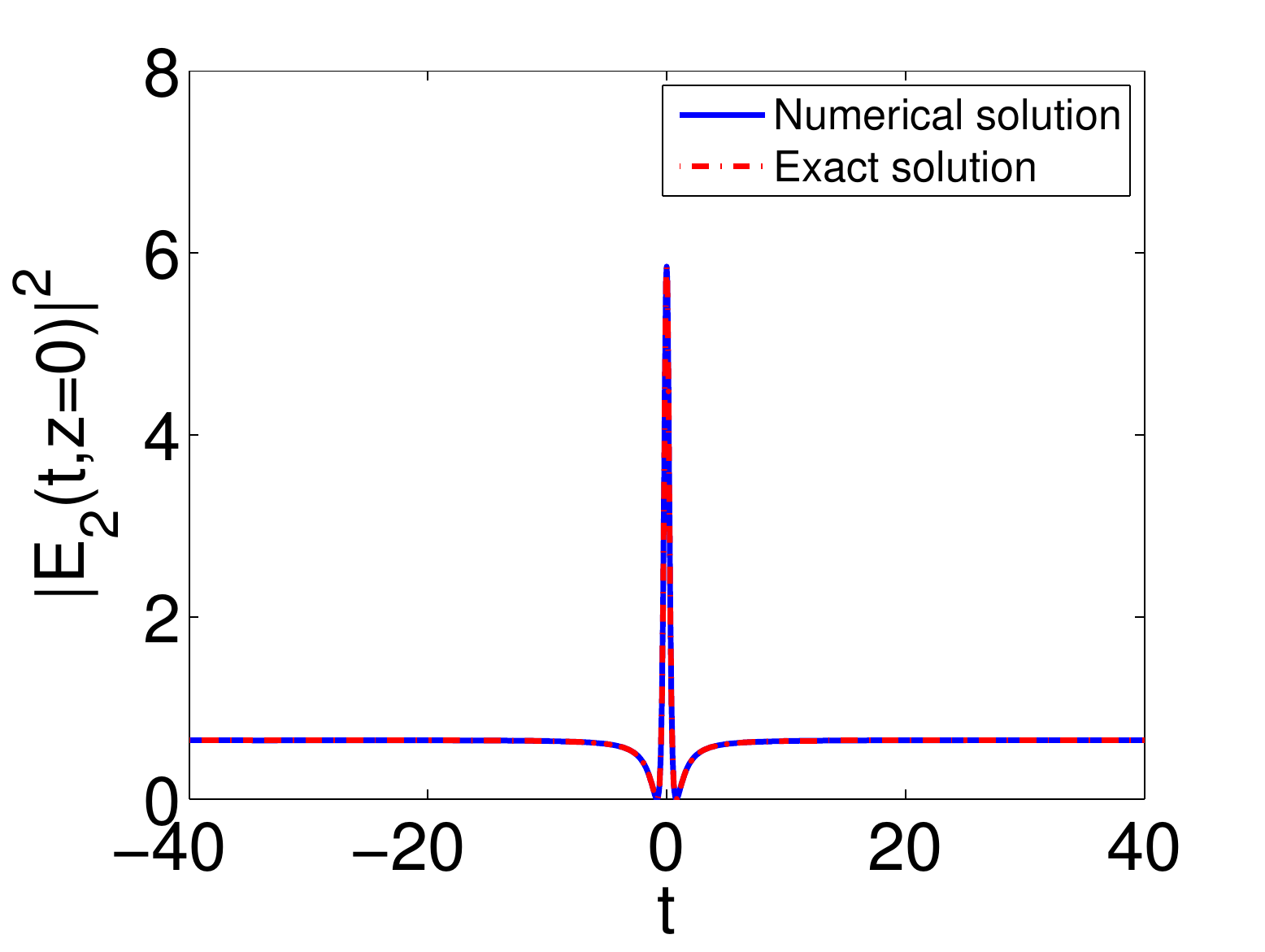}
\label{fig3b}
}
}
\mbox{\hspace{-0.2cm}
\subfigure[][]{\hspace{-1.0cm}
\includegraphics[height=.21\textheight, angle =0]{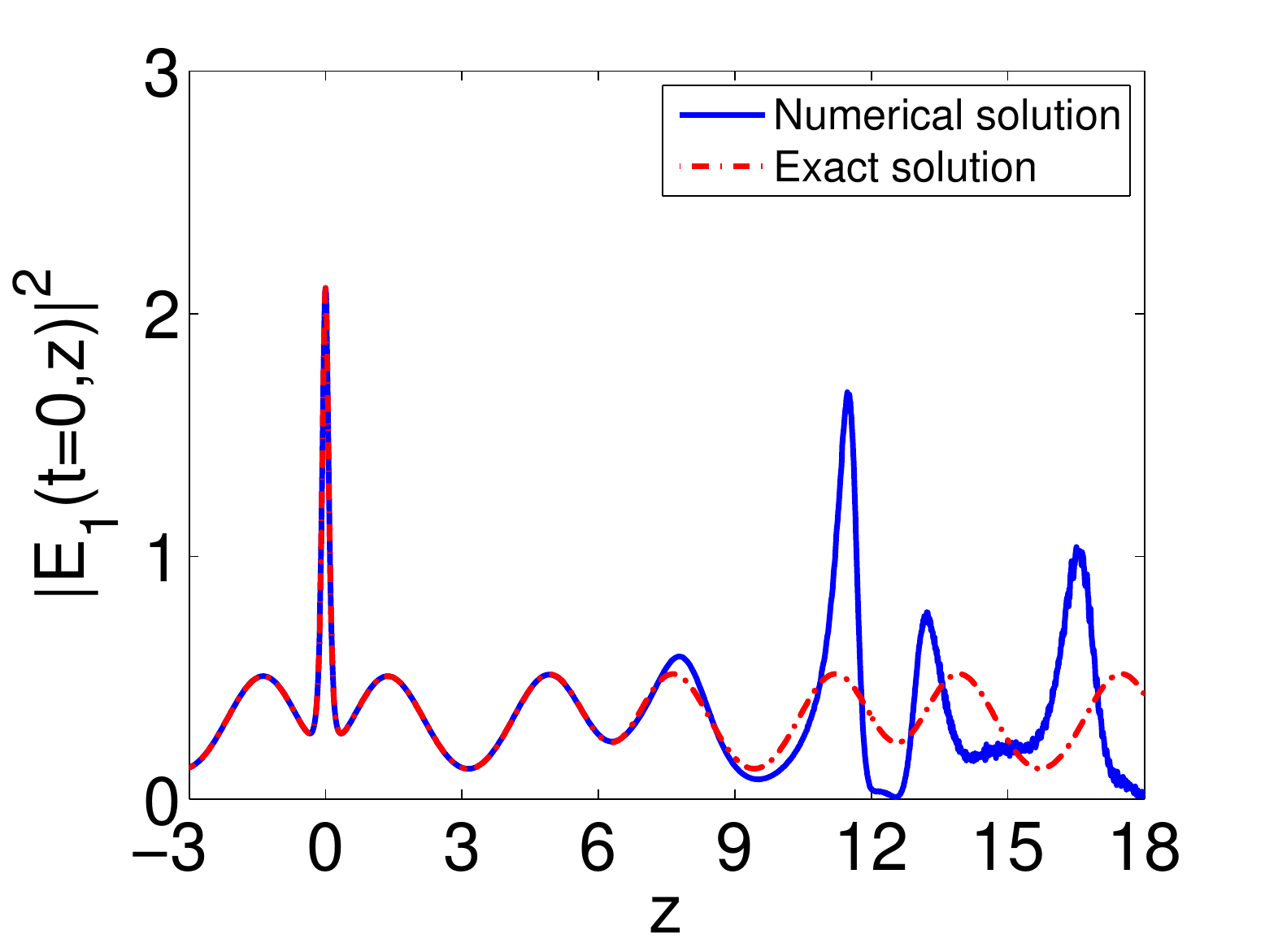}
\label{fig3c}
}
\subfigure[][]{\hspace{-0.2cm}
\includegraphics[height=.21\textheight, angle =0]{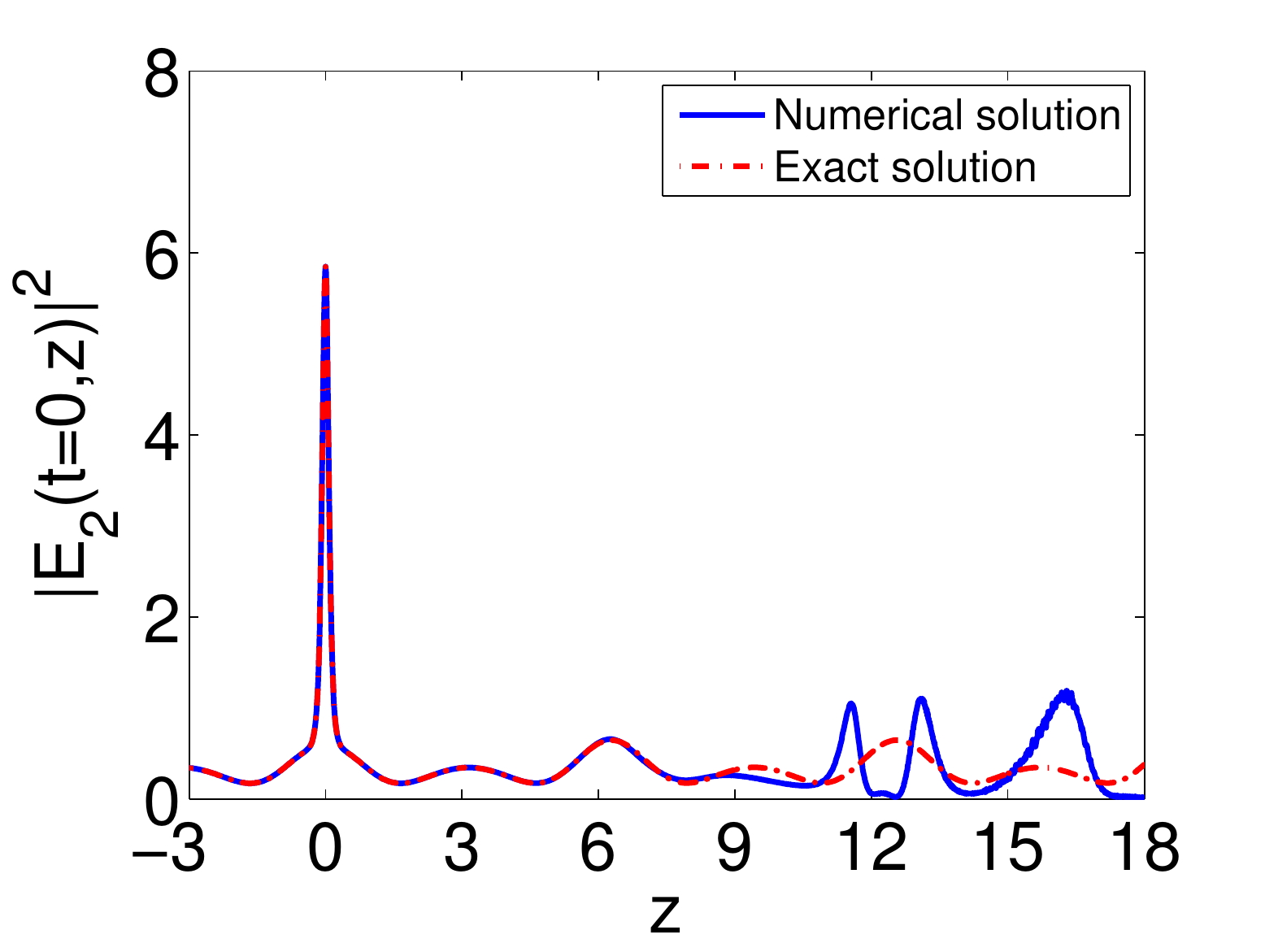}
\label{fig3d}
}
}
\mbox{\hspace{-0.2cm}
\subfigure[][]{\hspace{-1.0cm}
\includegraphics[height=.21\textheight, angle =0]{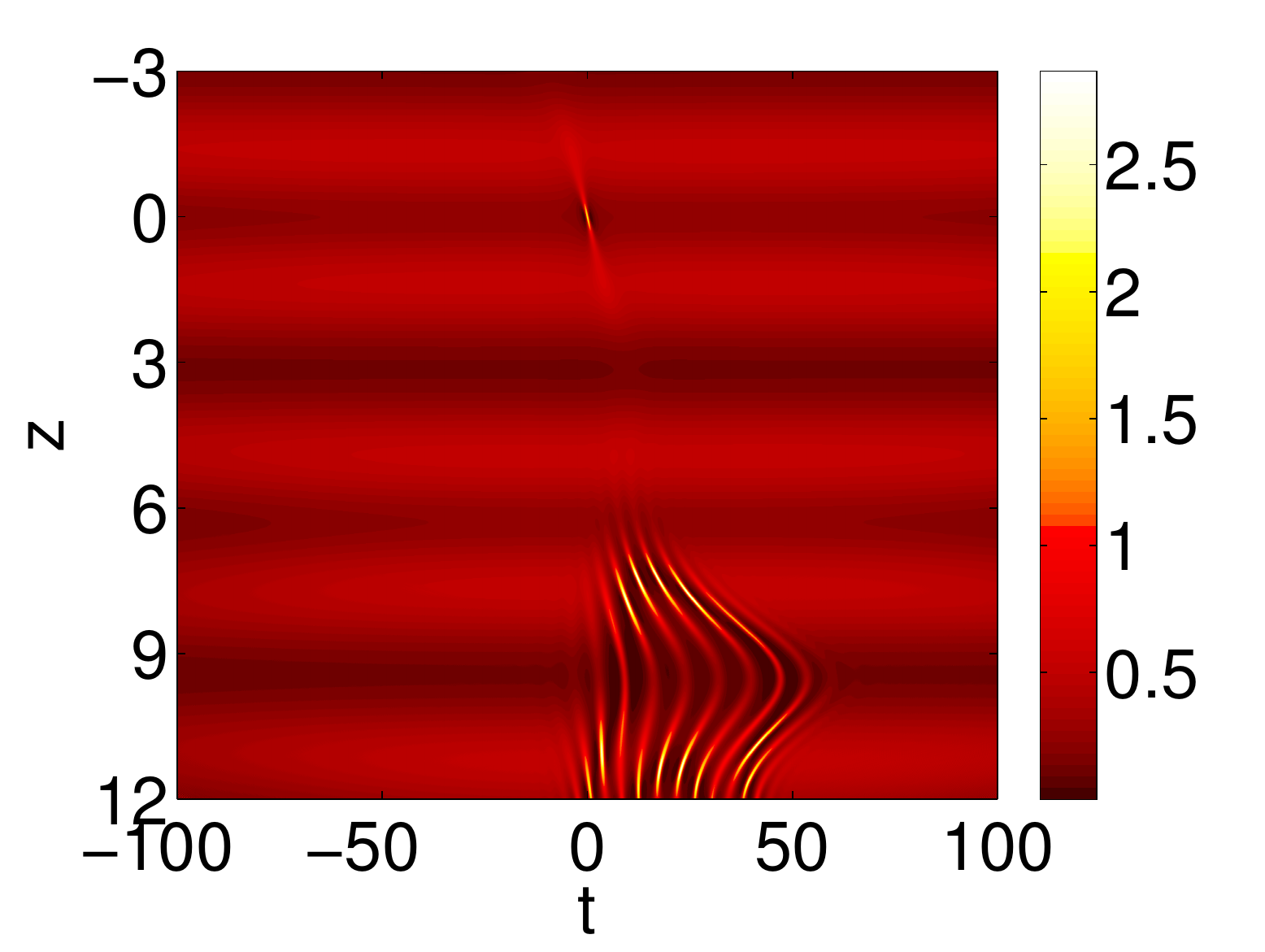}
\label{fig3e}
}
\subfigure[][]{\hspace{-0.2cm}
\includegraphics[height=.21\textheight, angle =0]{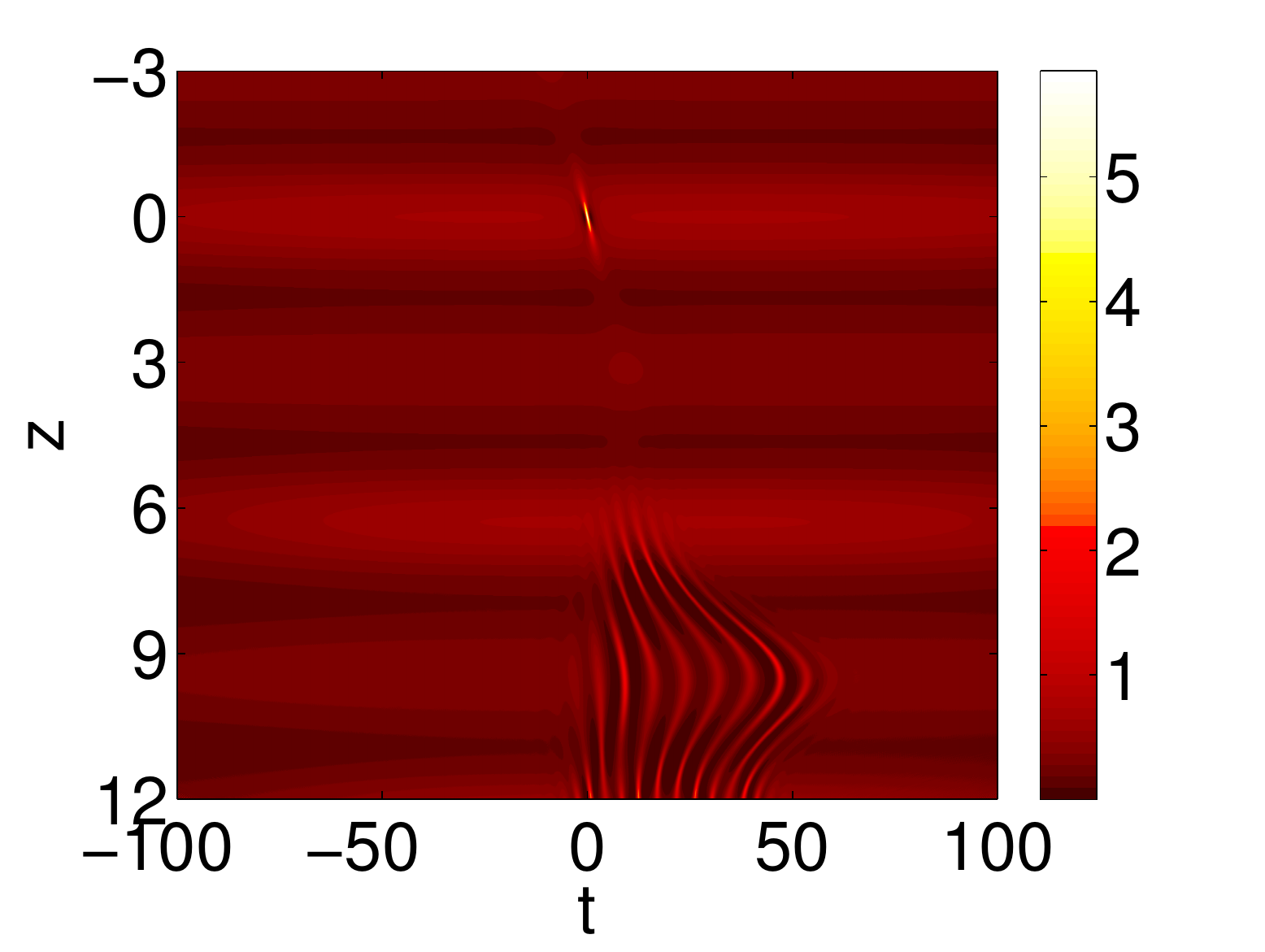}
\label{fig3f}
}
}
\mbox{\hspace{-0.2cm}
\subfigure[][]{\hspace{-1.0cm}
\includegraphics[height=.22\textheight, angle =0]{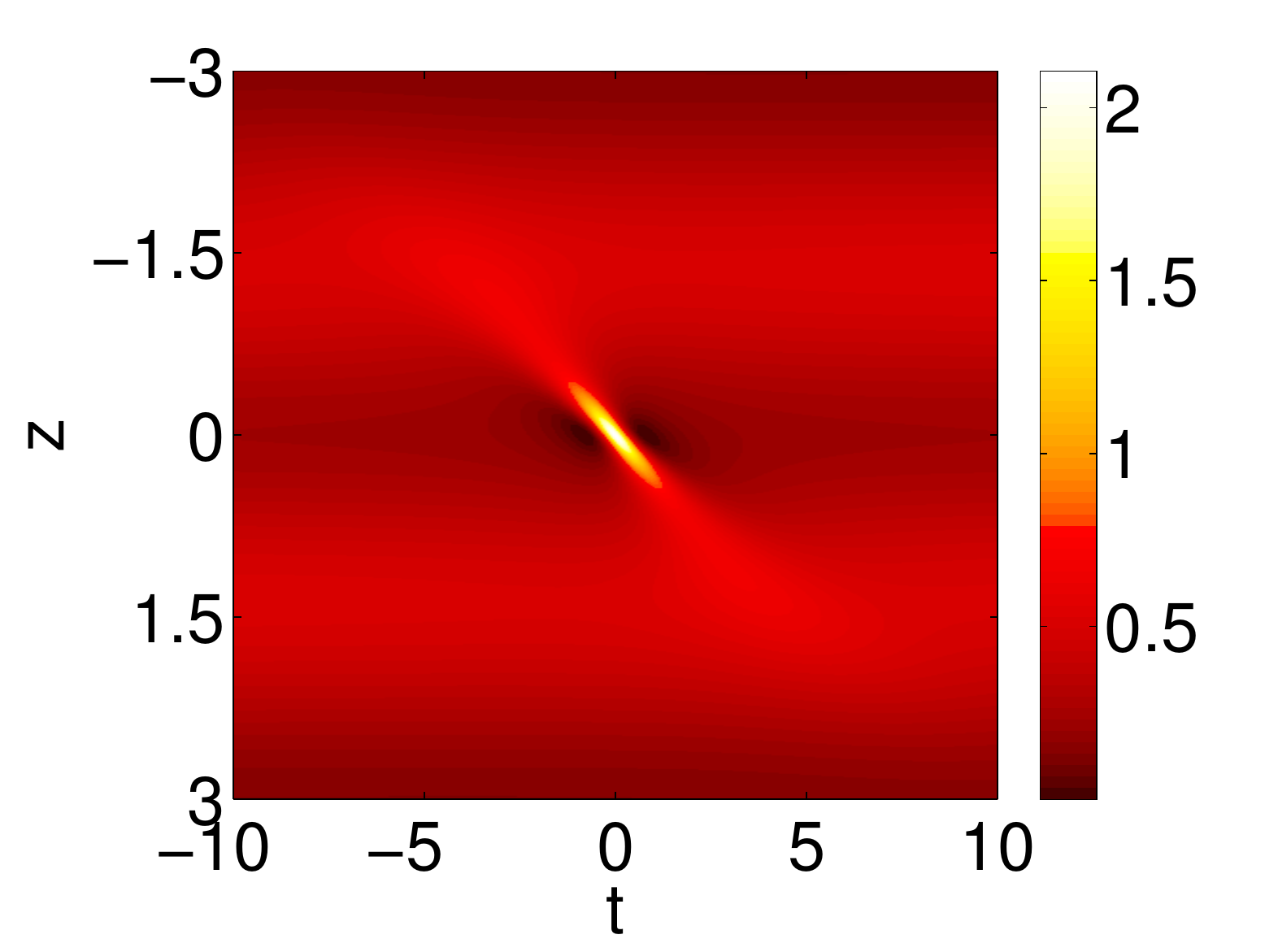}
\label{fig3g}
}
\subfigure[][]{\hspace{-0.2cm}
\includegraphics[height=.22\textheight, angle =0]{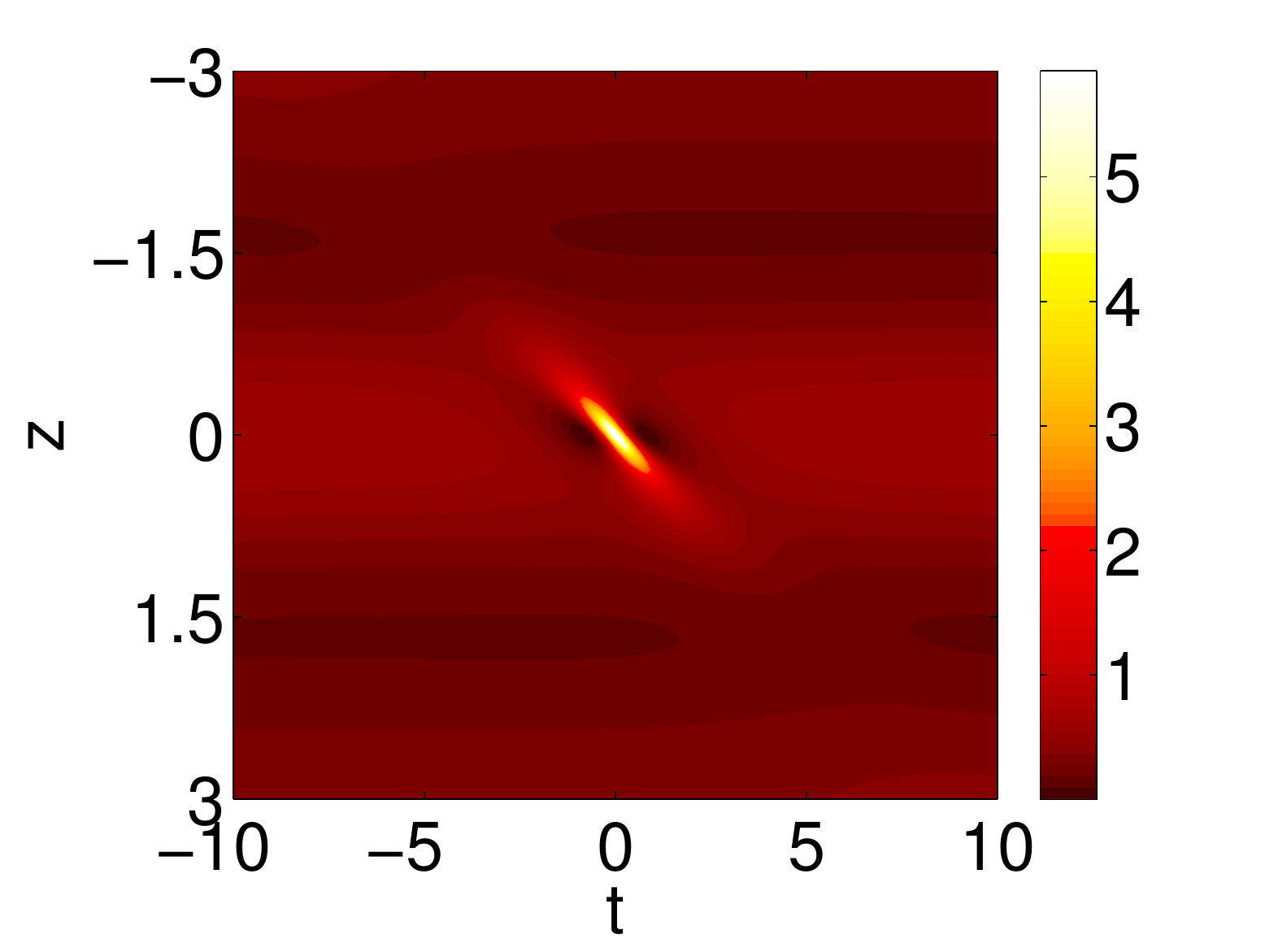}
\label{fig3h}
}
}
\end{center}
\caption{(Color online) Dynamics of rogue waves with a periodically modulated 
nonlinearity $\gamma(z)$ given by Eq.~(\ref{non-coefficient}) and the corresponding
trap frequency of Eq.~(\ref{trap}). Left and right panels correspond to density profiles
of the numerically obtained fields $E_{1}$ and $E_{2}$, respectively. The top panels (a-b)
show the spatial distribution of the intensities $|E_{j}|^{2}$ ($j=1,2$) evaluated at $z=0$,
whereas the second row panels (c-d) show their corresponding temporal evolution at $t=0$. 
The third row panels (e-f) show contour plots of the density profiles of the corresponding
rogue waves and the fourth row ones (g-h) zoom-ins of (e-f).}
\label{fig3}
\end{figure}

\begin{figure}[!pht]
\begin{center}
\vspace{-0.7cm}
\mbox{\hspace{-0.2cm}
\subfigure[][]{\hspace{-1.0cm}
\includegraphics[height=.21\textheight, angle =0]{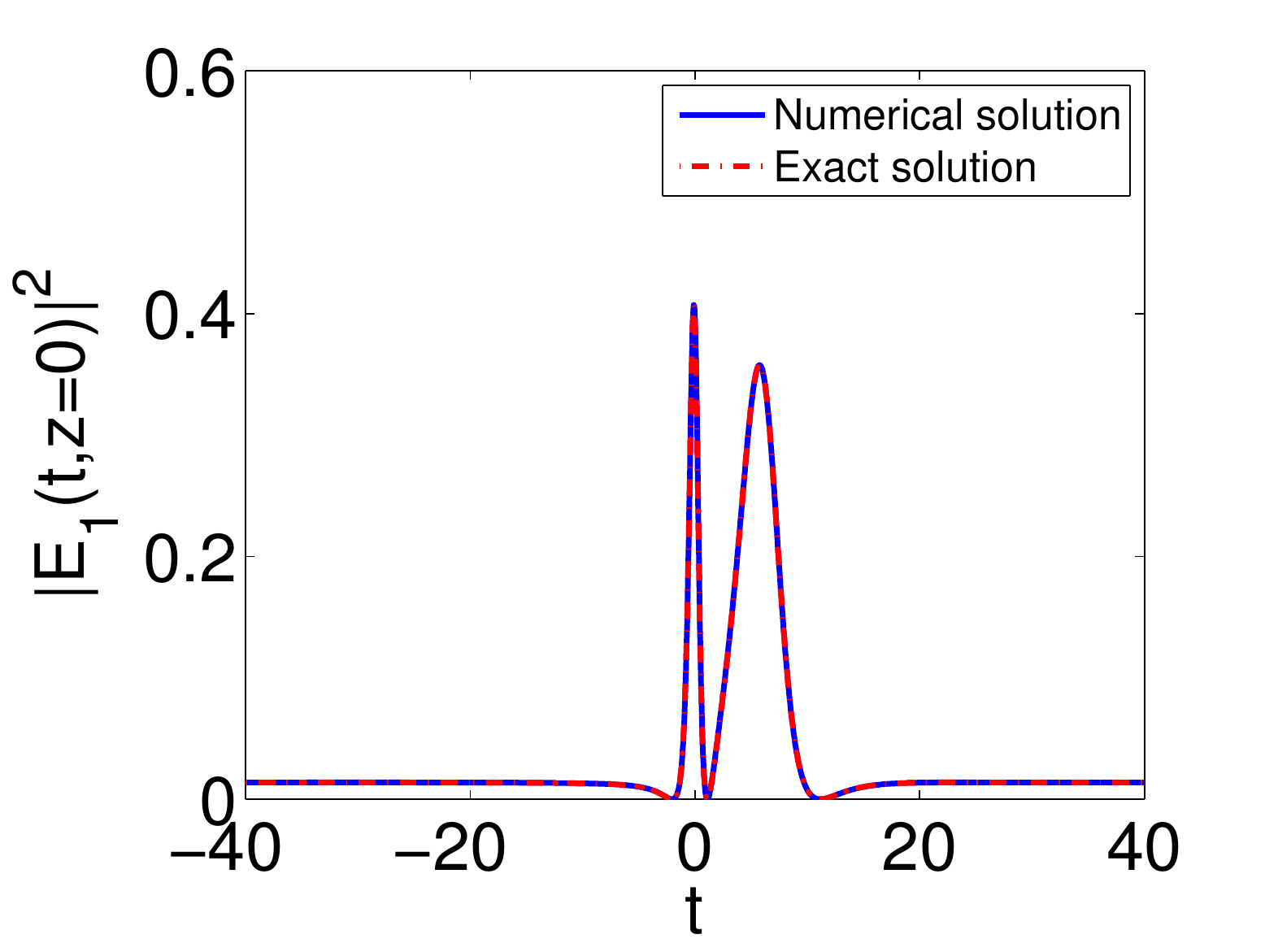}
\label{fig4a}
}
\subfigure[][]{\hspace{-0.2cm}
\includegraphics[height=.21\textheight, angle =0]{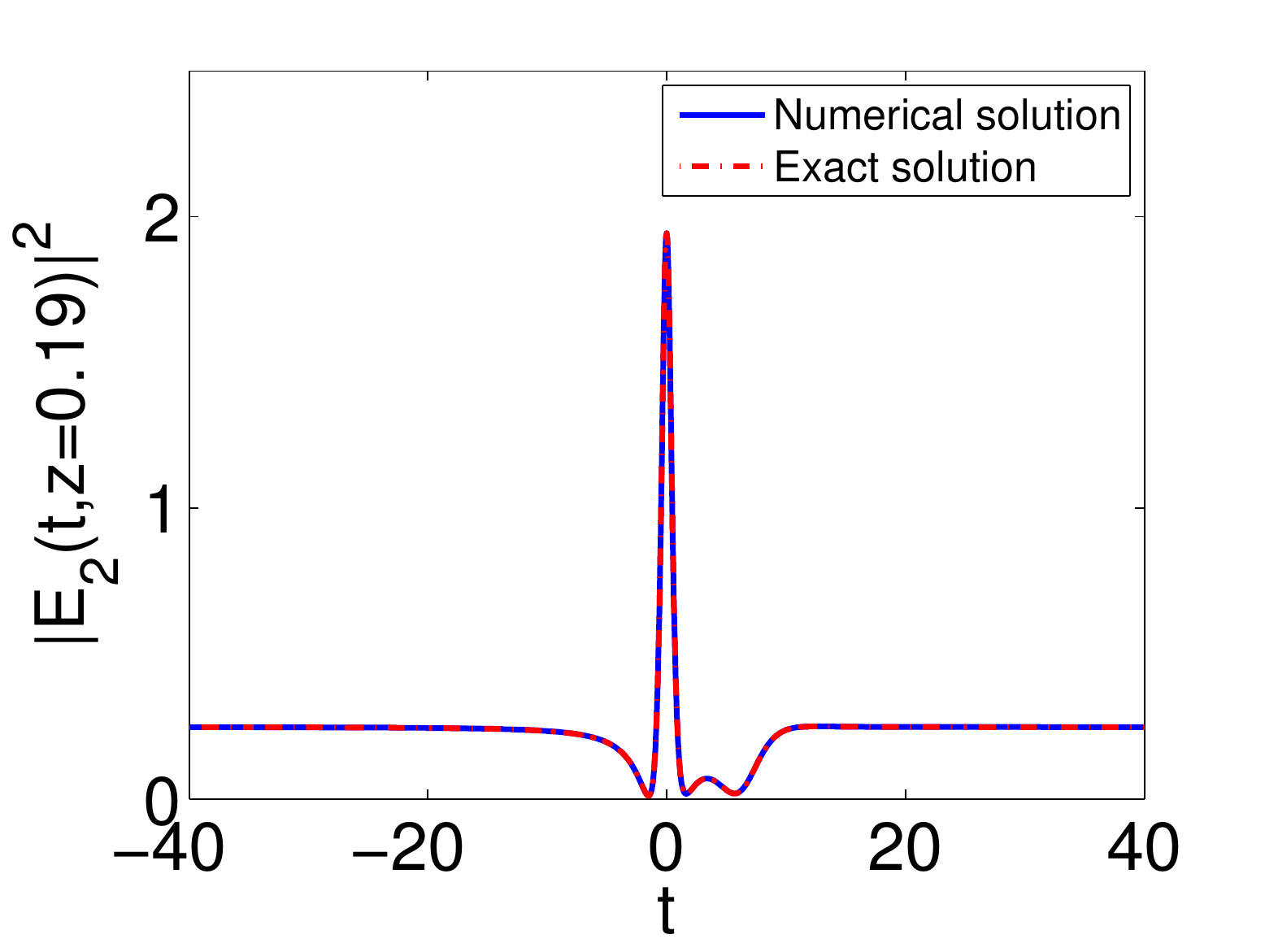}
\label{fig4b}
}
}
\mbox{\hspace{-0.2cm}
\subfigure[][]{\hspace{-1.0cm}
\includegraphics[height=.21\textheight, angle =0]{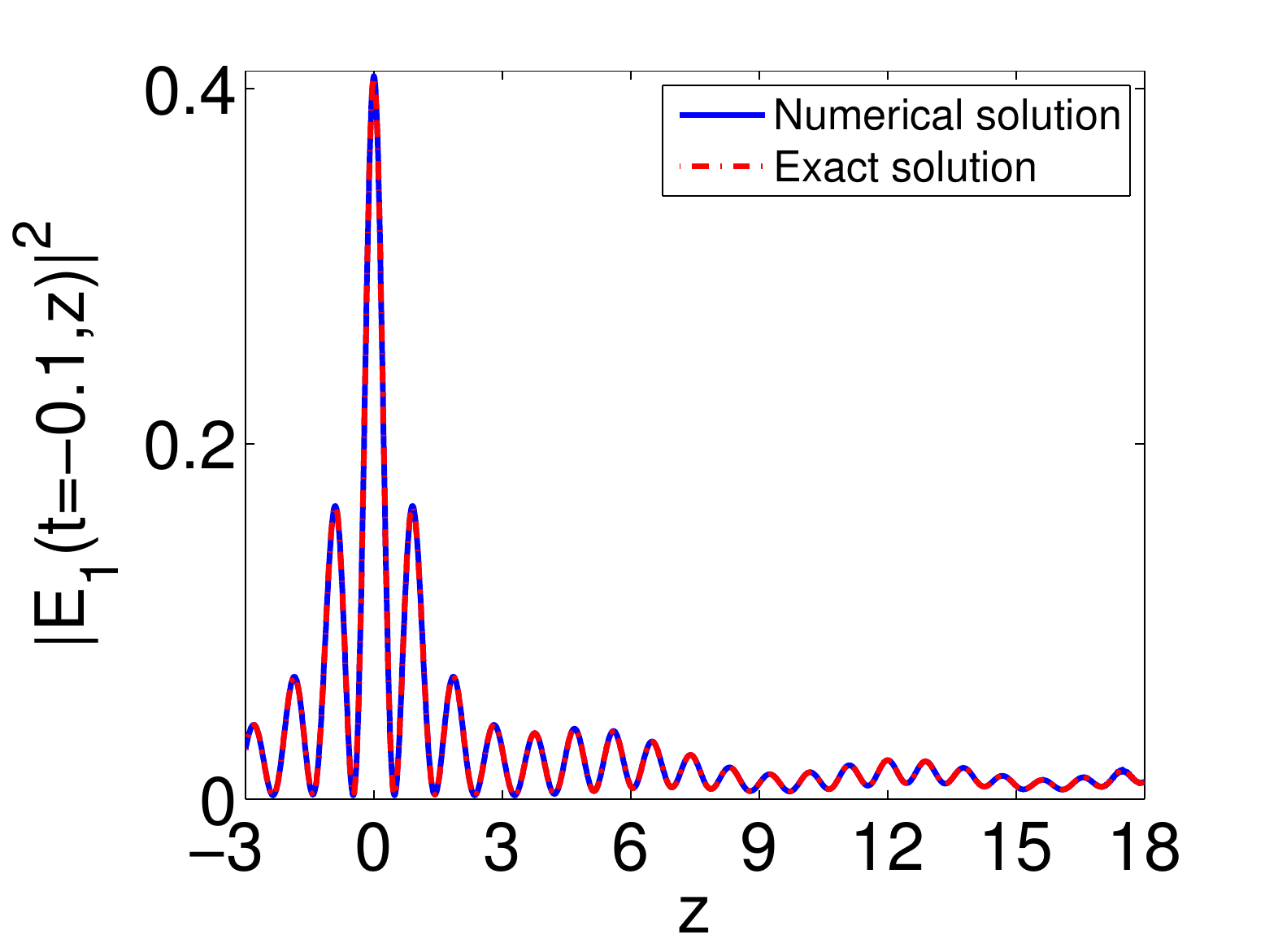}
\label{fig4c}
}
\subfigure[][]{\hspace{-0.2cm}
\includegraphics[height=.21\textheight, angle =0]{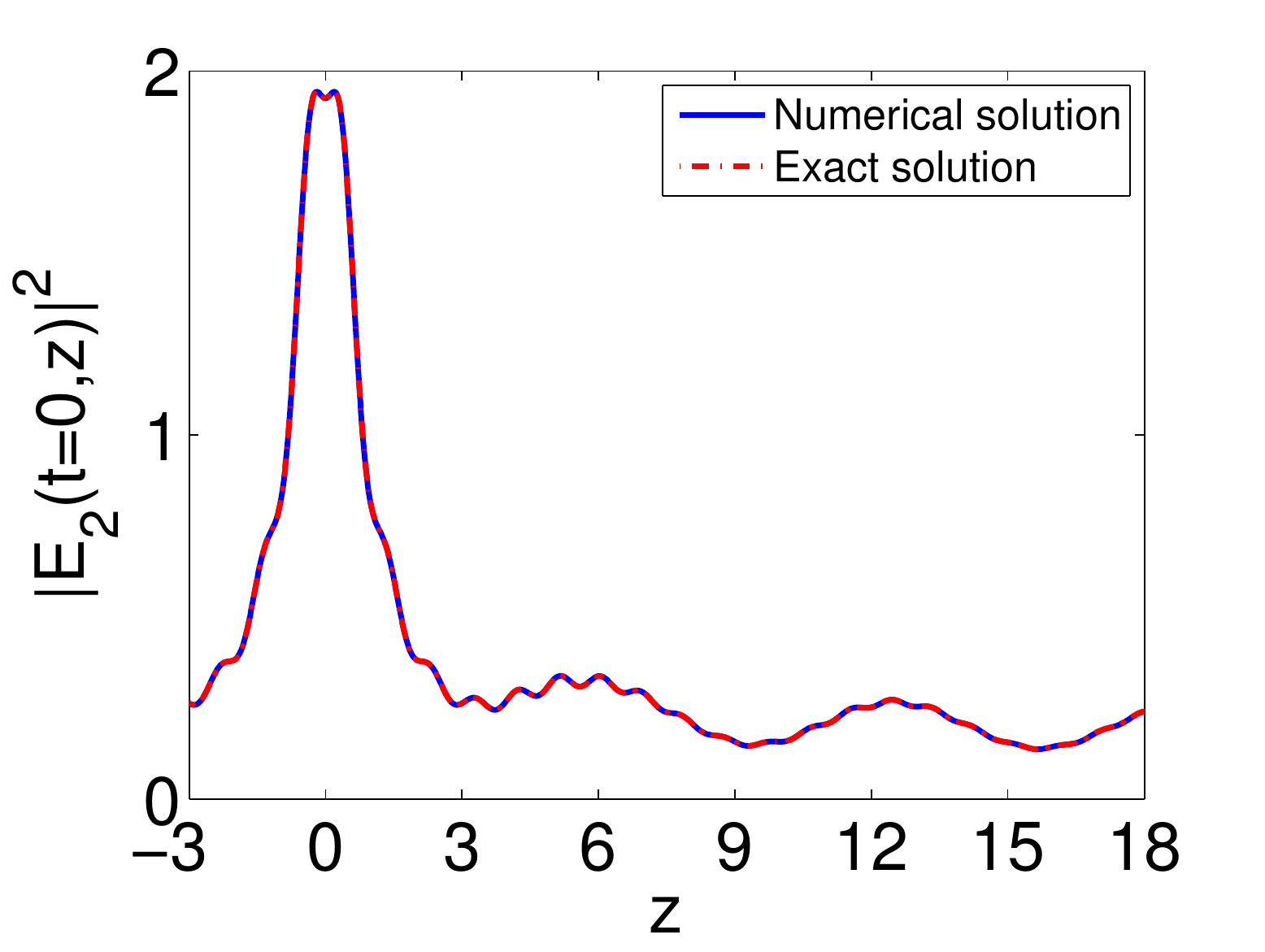}
\label{fig4d}
}
}
\mbox{\hspace{-0.2cm}
\subfigure[][]{\hspace{-1.0cm}
\includegraphics[height=.21\textheight, angle =0]{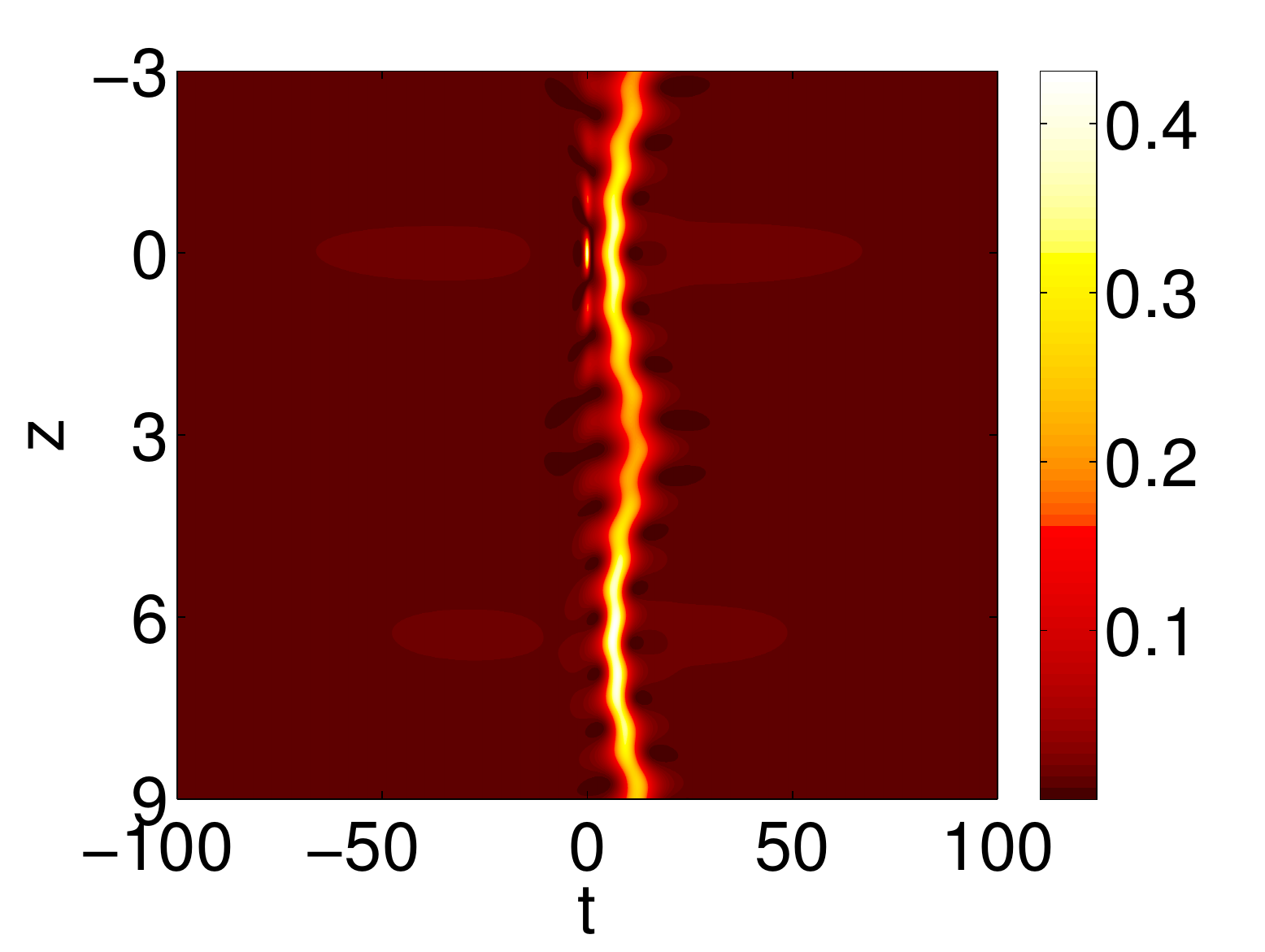}
\label{fig4e}
}
\subfigure[][]{\hspace{-0.2cm}
\includegraphics[height=.21\textheight, angle =0]{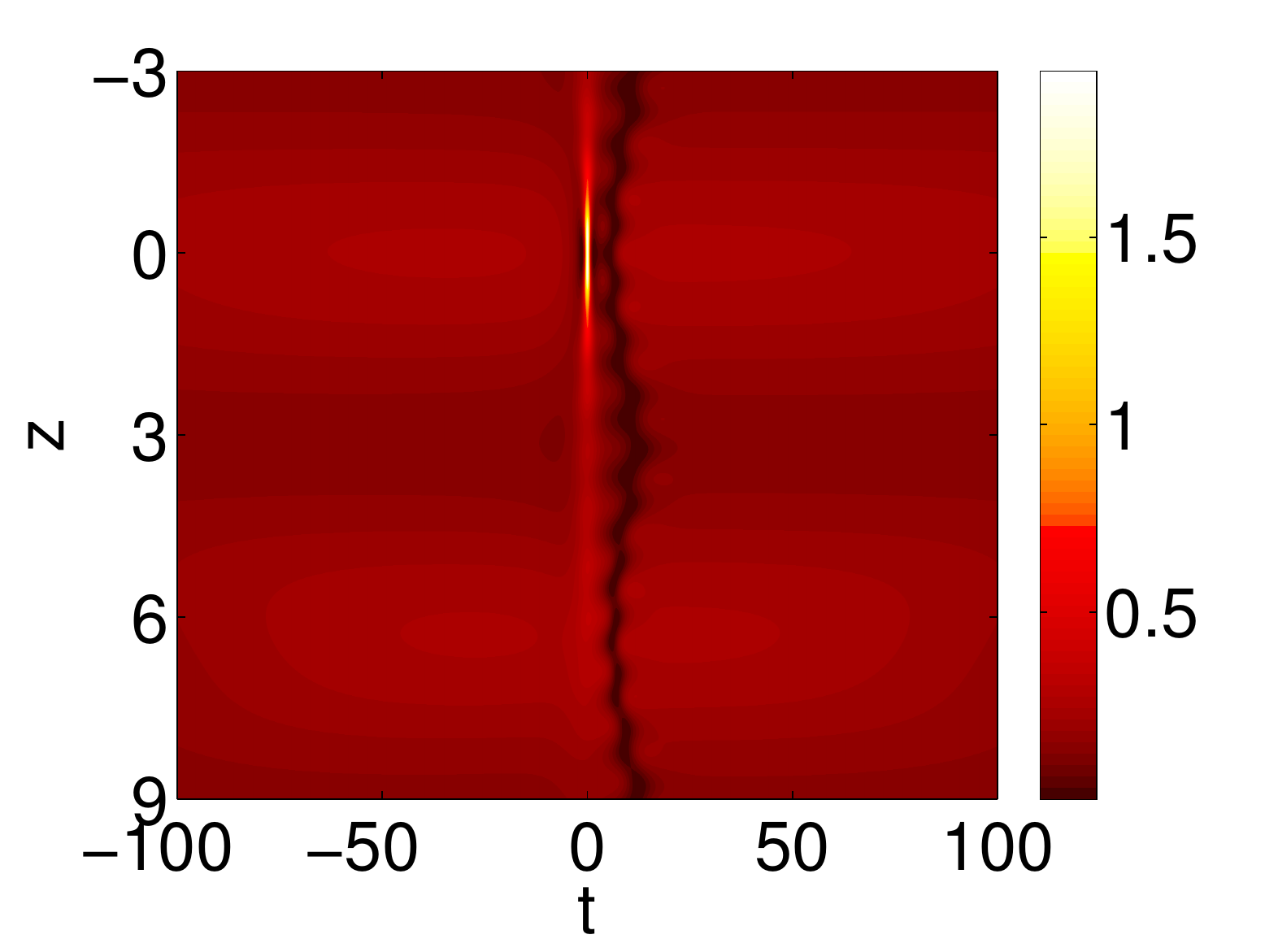}
\label{fig4f}
}
}
\mbox{\hspace{-0.2cm}
\subfigure[][]{\hspace{-1.0cm}
\includegraphics[height=.22\textheight, angle =0]{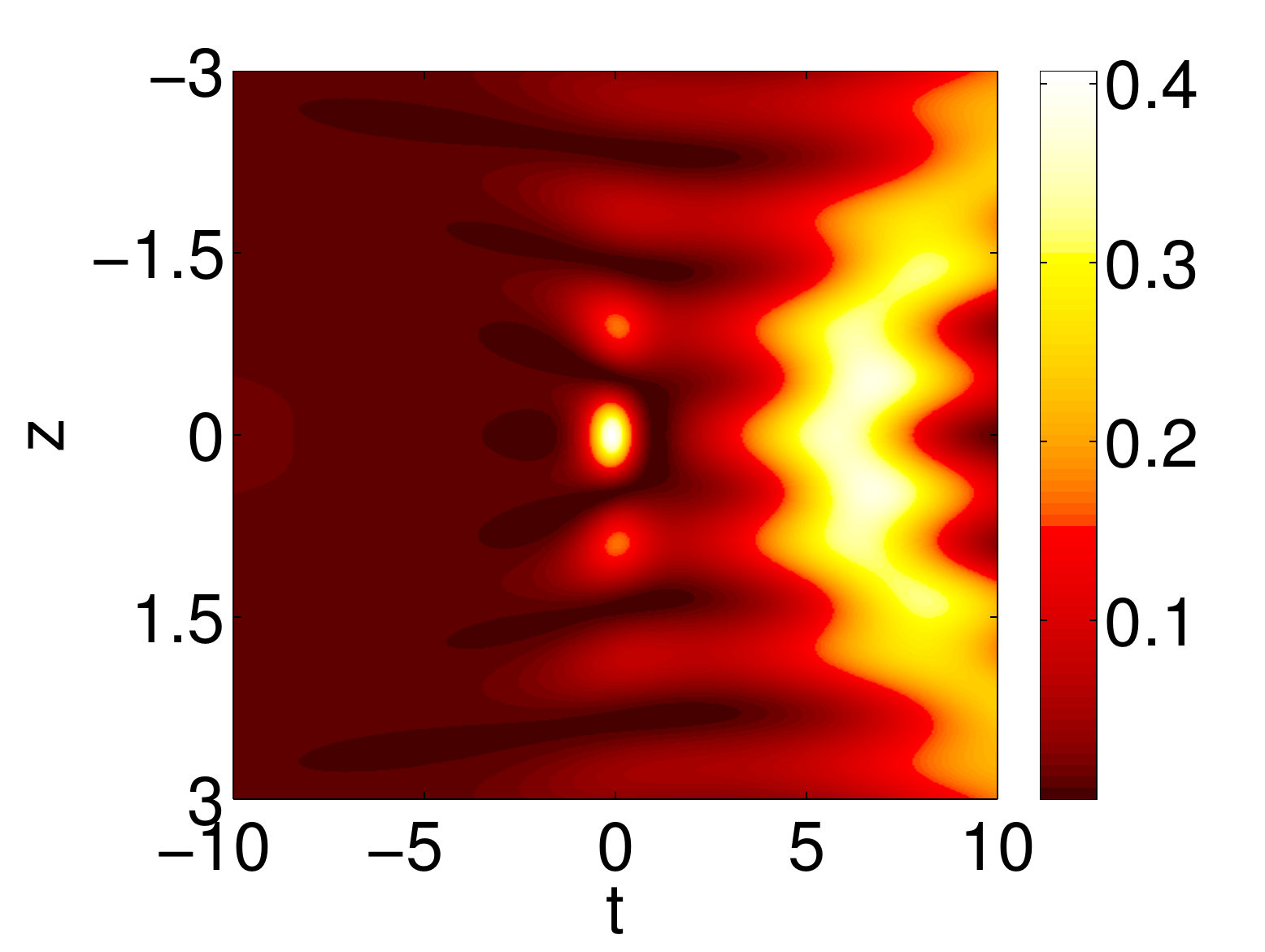}
\label{fig4g}
}
\subfigure[][]{\hspace{-0.2cm}
\includegraphics[height=.22\textheight, angle =0]{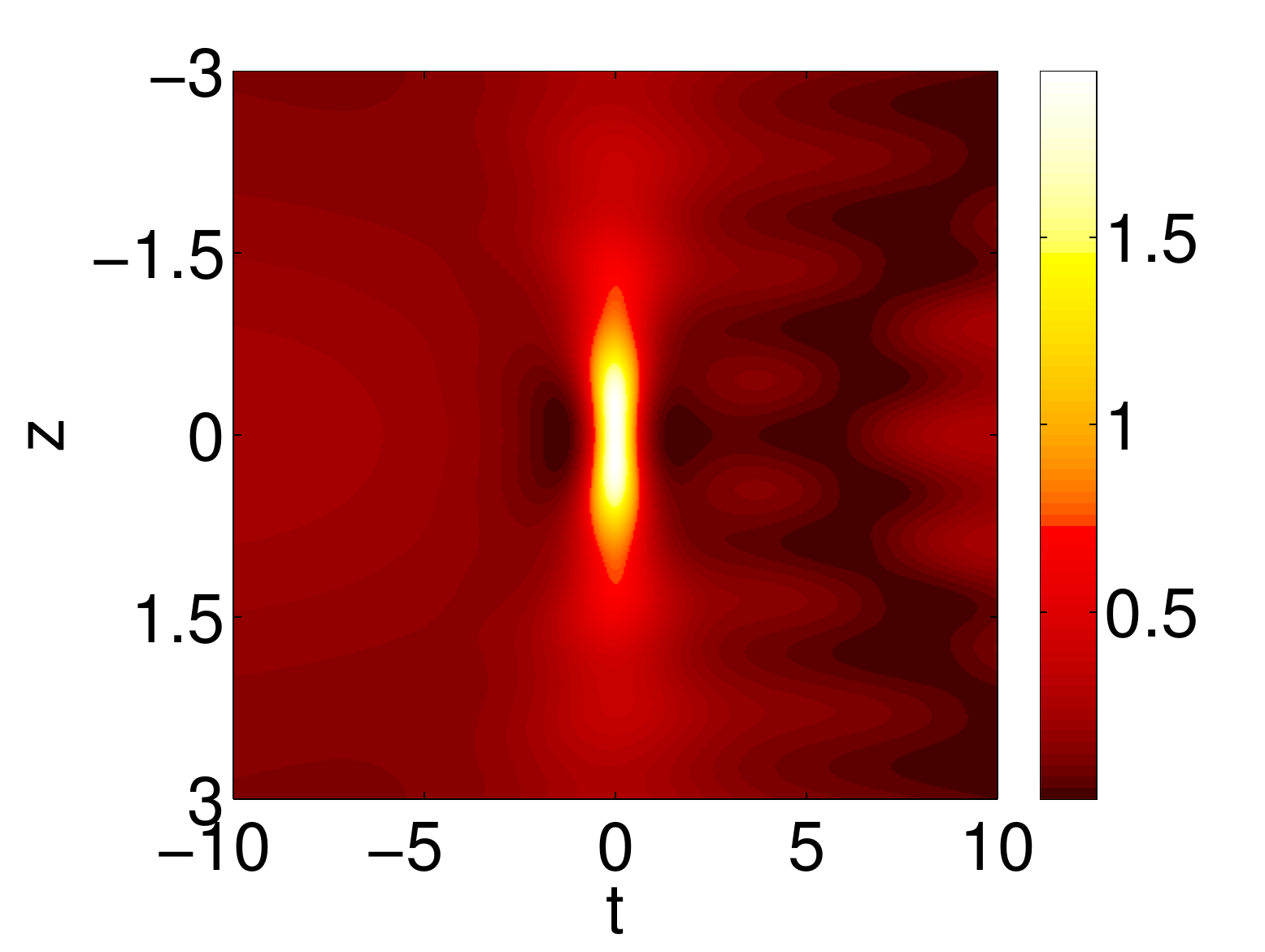}
\label{fig4h}
}
}
\end{center}
\caption{(Color online) Dynamics of co-existing Peregrine and DB boomeronic solitons 
with a periodically modulated nonlinearity~(\ref{non-coefficient}). Left and right 
panels correspond to density profiles of the numerically obtained fields $E_{1}$ and
$E_{2}$, respectively. The top row panels (a-b) show the spatial distribution of the
intensities $|E_{1}|^{2}$ at $z=0$ (left) and $|E_{2}|^{2}$ at $z=0.19$ (right), whereas
the second row panels (c-d) show their corresponding temporal evolution at $t=-0.1$ and
$t=0$, respectively. Again, the panels (e-h) show contour plots of the density profiles
of the corresponding waves.}
\label{fig4}
\end{figure}

\begin{figure}[!pht]                              
\begin{center}
\vspace{-0.7cm}
\mbox{\hspace{-0.2cm}
\subfigure[][]{\hspace{-1.0cm}                    
\includegraphics[height=.21\textheight, angle =0]{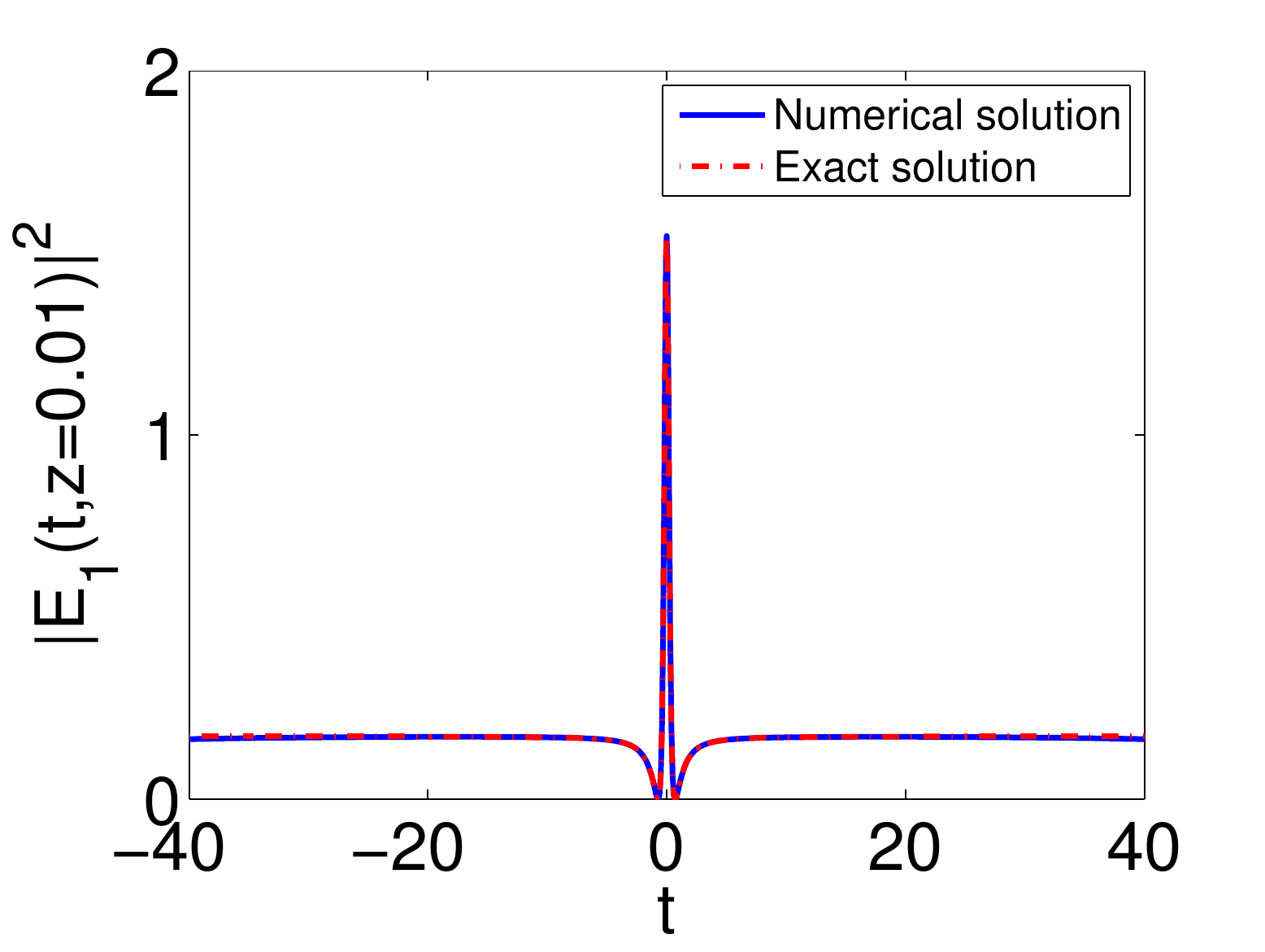}
\label{fig5a}
}
\subfigure[][]{\hspace{-0.2cm}
\includegraphics[height=.21\textheight, angle =0]{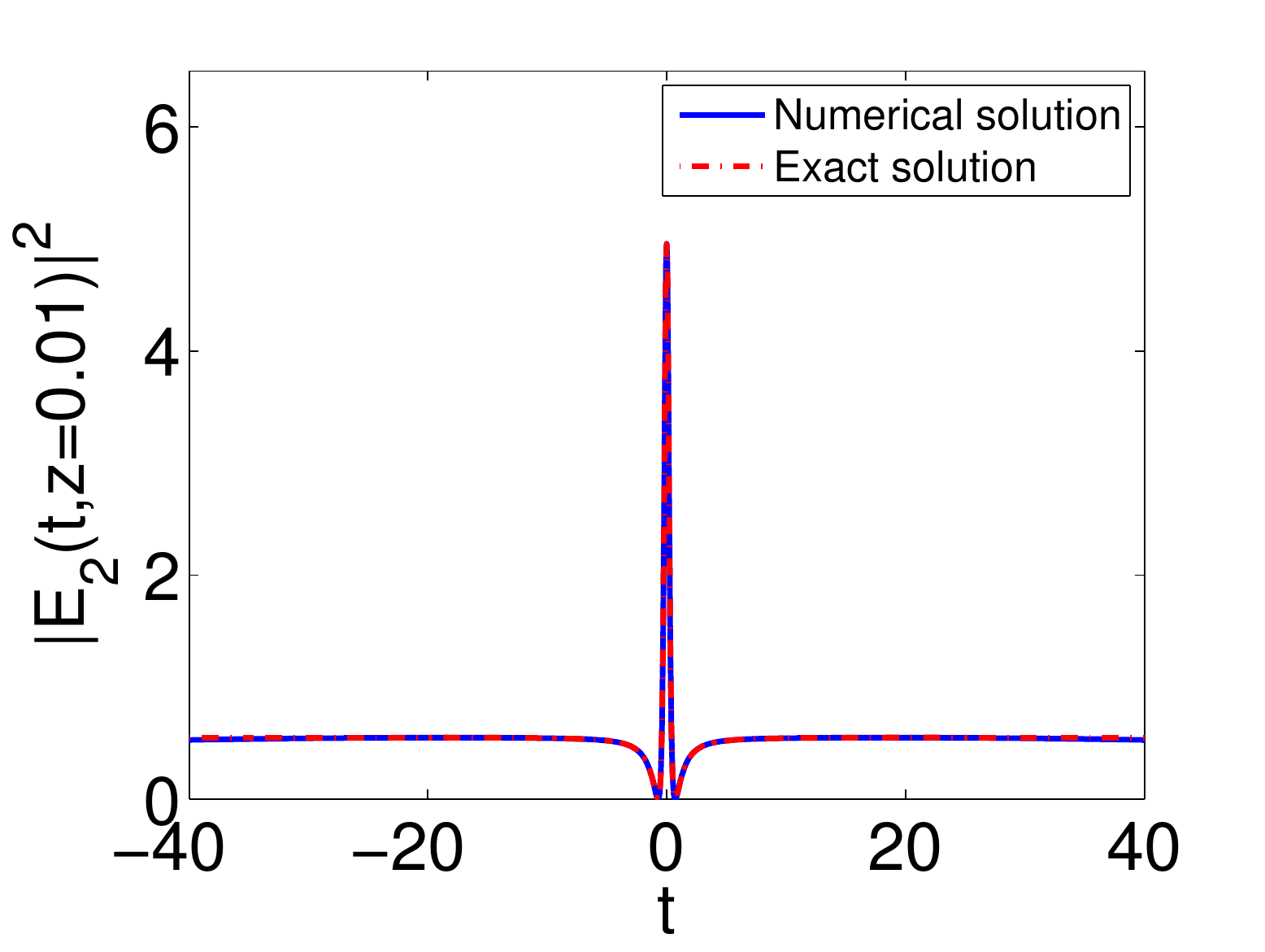}
\label{fig5b}
}
}
\mbox{\hspace{-0.2cm}
\subfigure[][]{\hspace{-1.0cm}
\includegraphics[height=.21\textheight, angle =0]{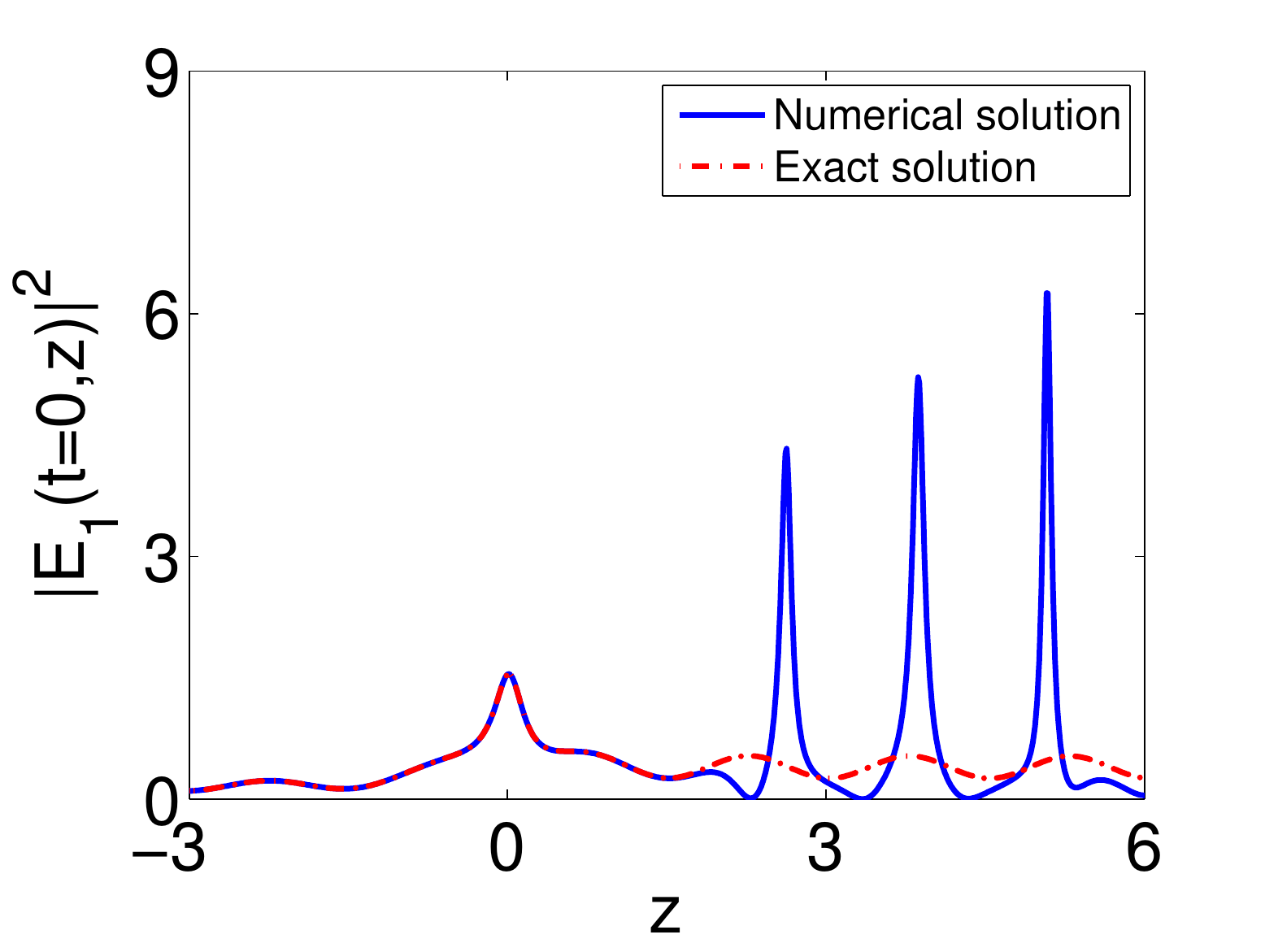}
\label{fig5c}
}
\subfigure[][]{\hspace{-0.2cm}
\includegraphics[height=.21\textheight, angle =0]{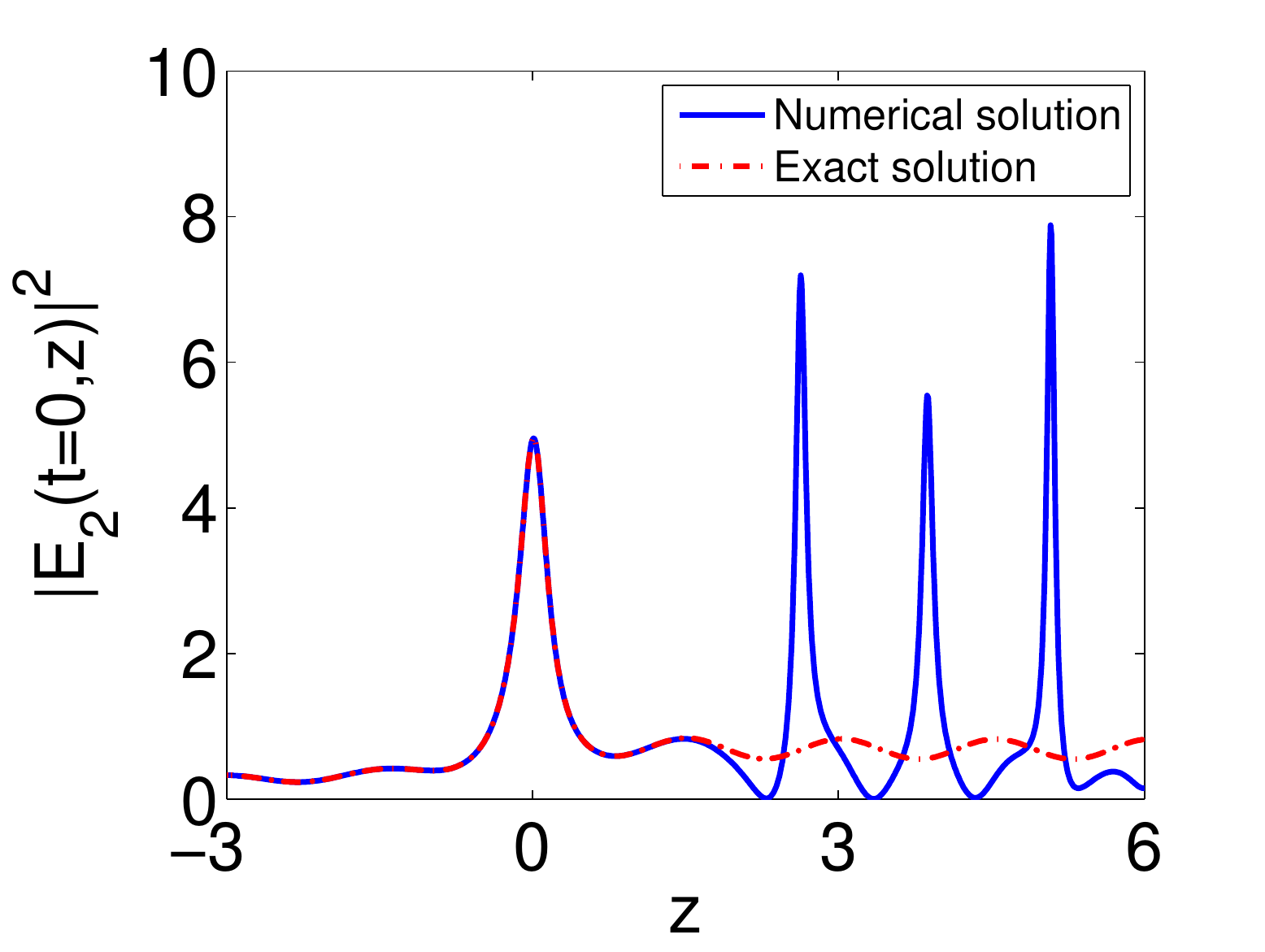}
\label{fig5d}
}
}
\mbox{\hspace{-0.2cm}
\subfigure[][]{\hspace{-1.0cm}
\includegraphics[height=.21\textheight, angle =0]{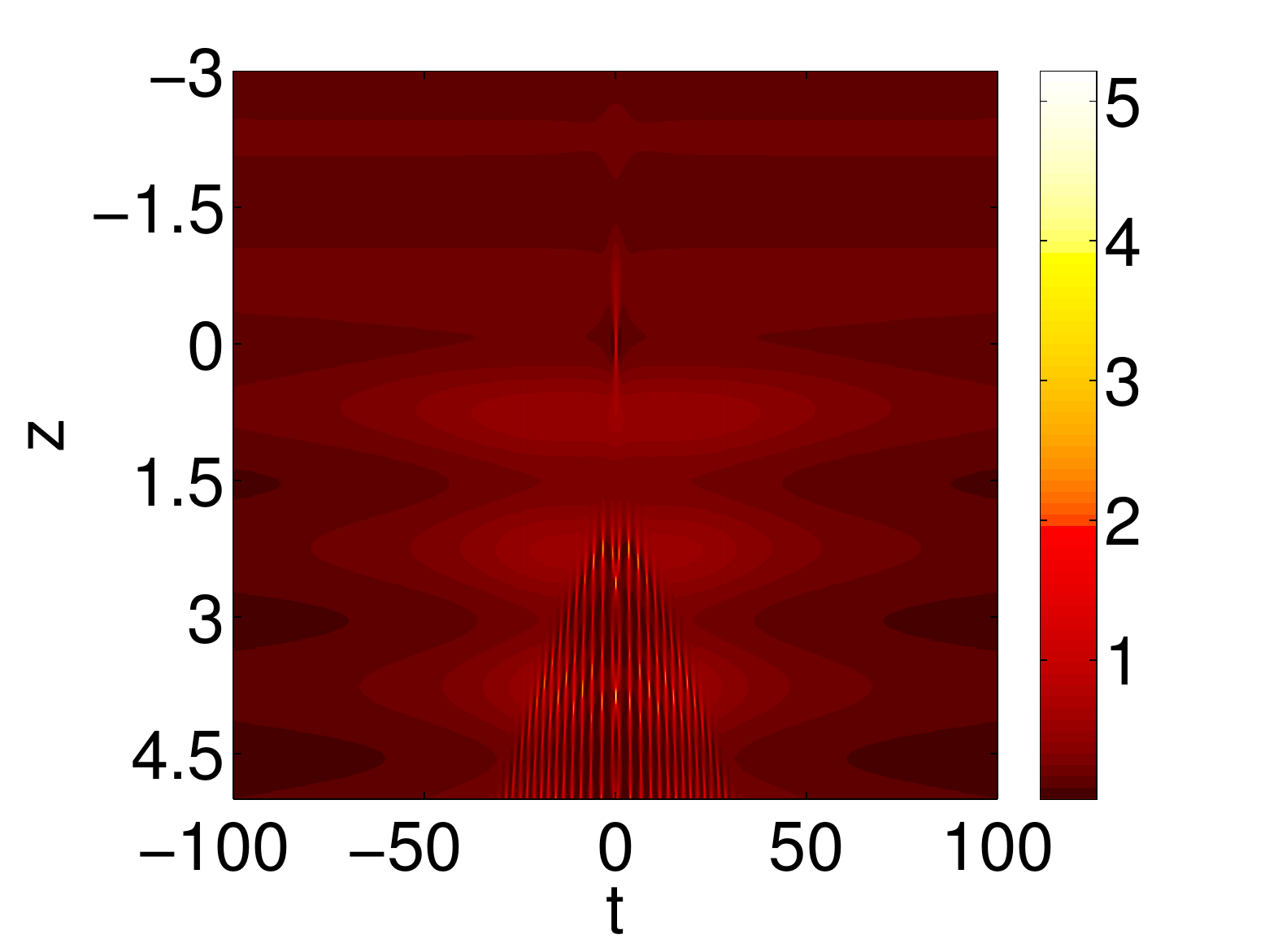}
\label{fig5e}
}
\subfigure[][]{\hspace{-0.2cm}
\includegraphics[height=.21\textheight, angle =0]{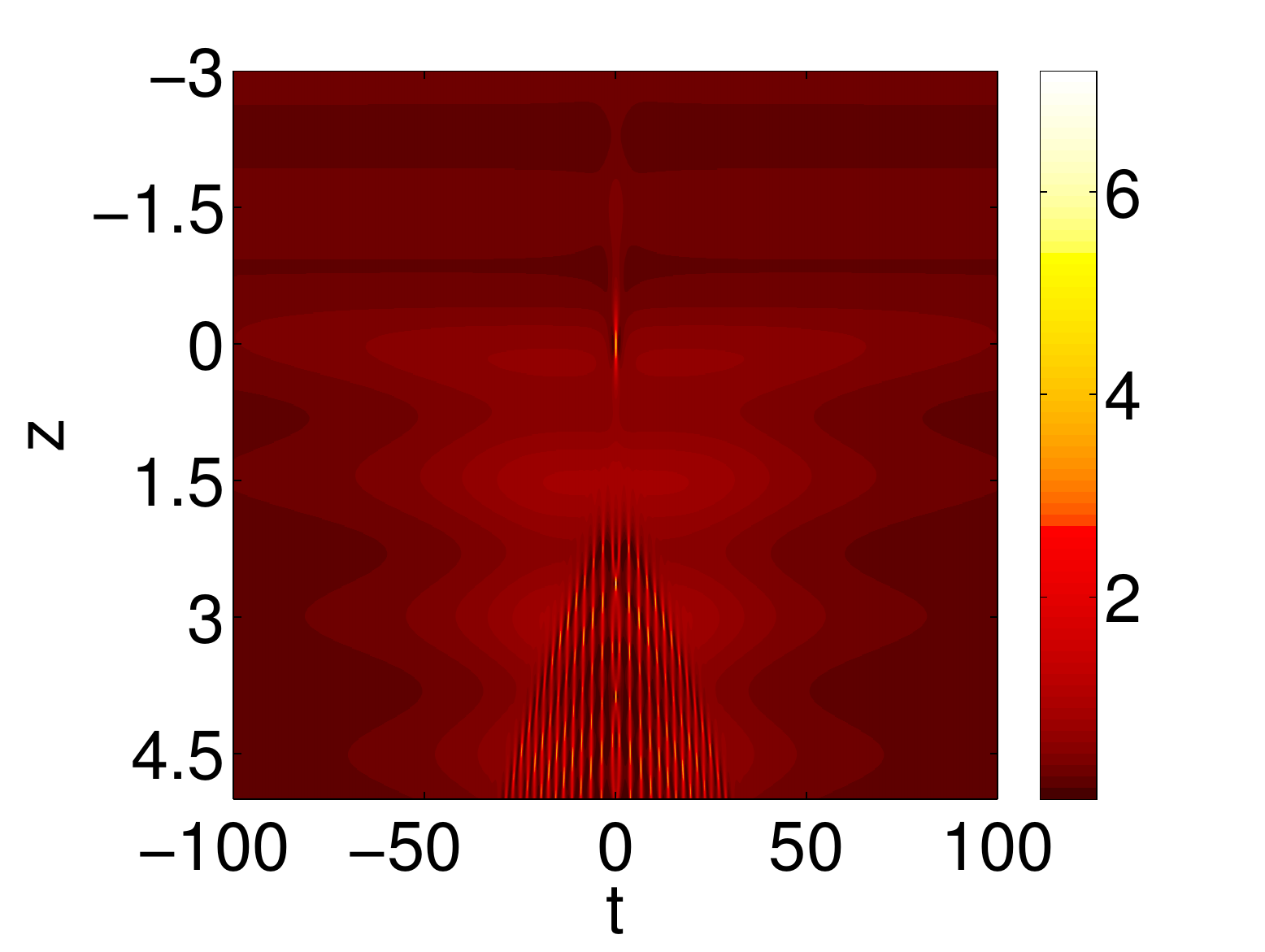}
\label{fig5f}
}
}
\mbox{\hspace{-0.2cm}
\subfigure[][]{\hspace{-1.0cm}
\includegraphics[height=.21\textheight, angle =0]{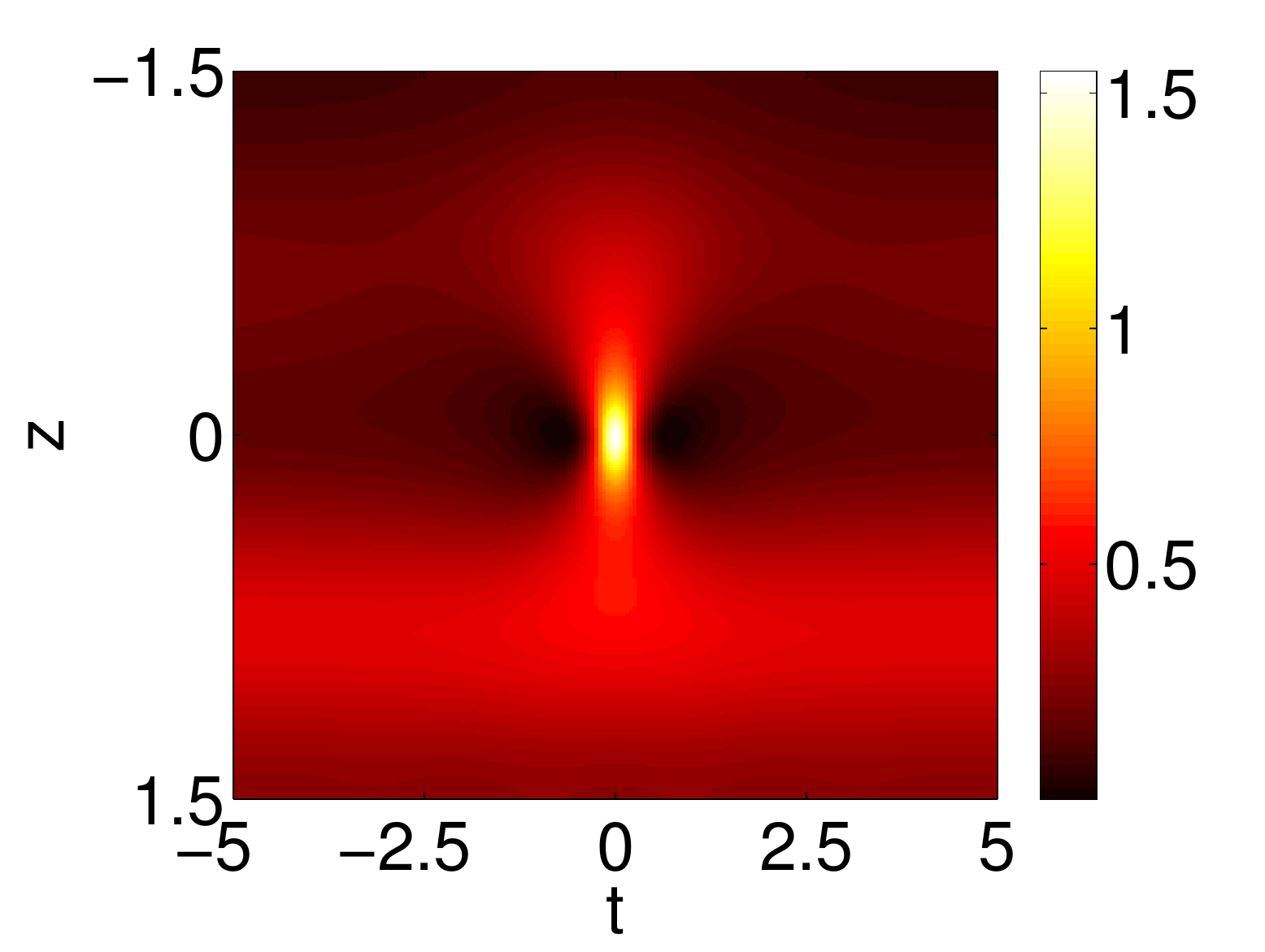}
\label{fig5g}
}
\subfigure[][]{\hspace{-0.2cm}
\includegraphics[height=.21\textheight, angle =0]{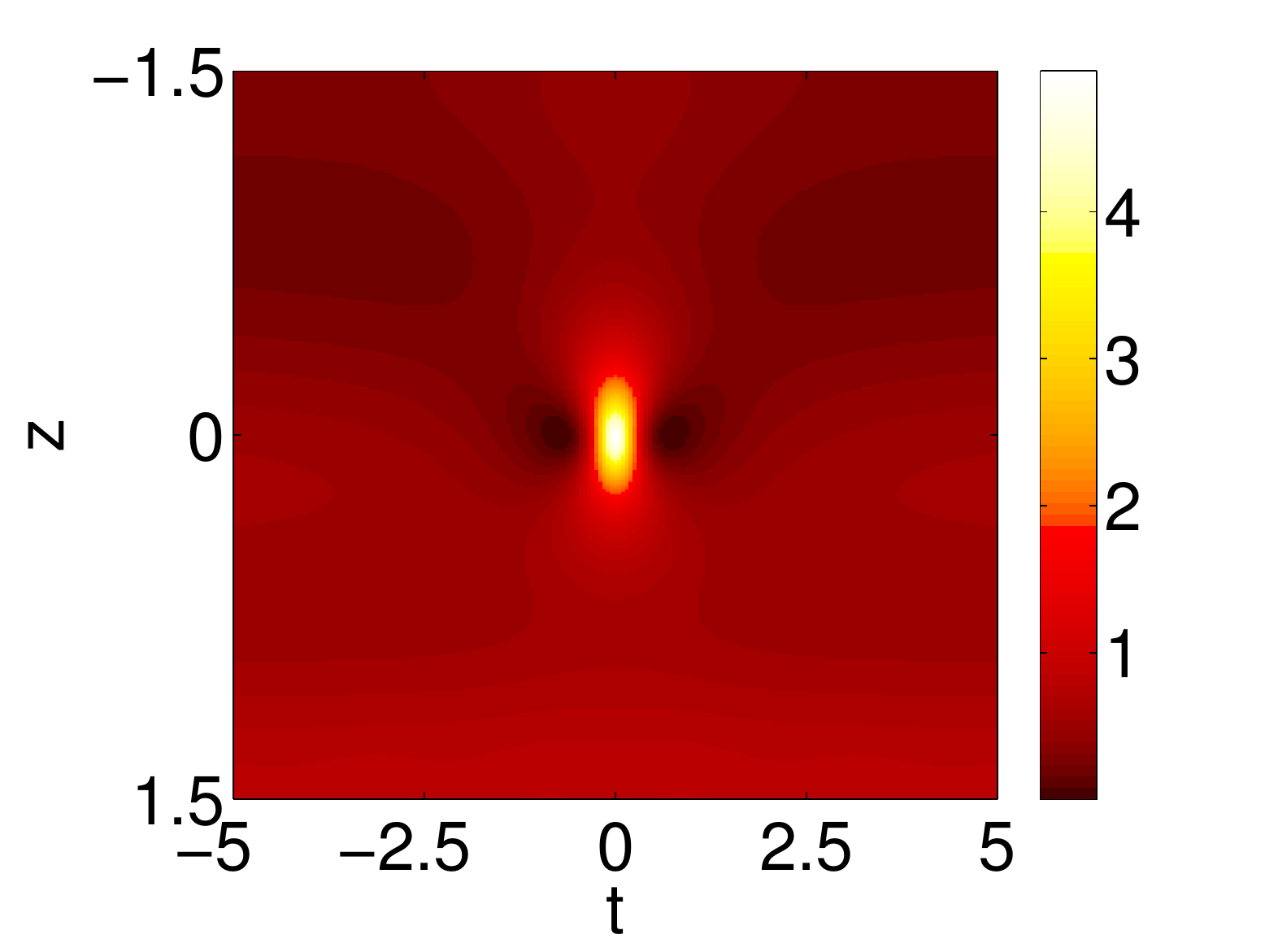}
\label{fig5h}
}
}
\end{center}
\caption{(Color online) Same as Fig.~\ref{fig3} but for the kink-like modulated nonlinearity
given by  Eq.~(\ref{eq12a}) and the trap frequency of Eq.~(\ref{eq12b}). Left and right 
panels correspond to density profiles of the numerically obtained fields $E_{1}$ and $E_{2}$,
respectively. The top row panels (a-b) show the spatial distribution of the intensities 
$|E_{j}|^{2}$ ($j=1,2$) evaluated at $z=0.01$, whereas the second row panels (c-d) show 
their corresponding temporal evolution at $t=0$. The panels (e-h) show contour plots of 
the density profiles of the corresponding rogue waves.}
\label{fig5}
\end{figure}

\begin{figure}[!pht]
\begin{center}
\vspace{-0.7cm}
\mbox{\hspace{-0.2cm}
\subfigure[][]{\hspace{-1.0cm}
\includegraphics[height=.21\textheight, angle =0]{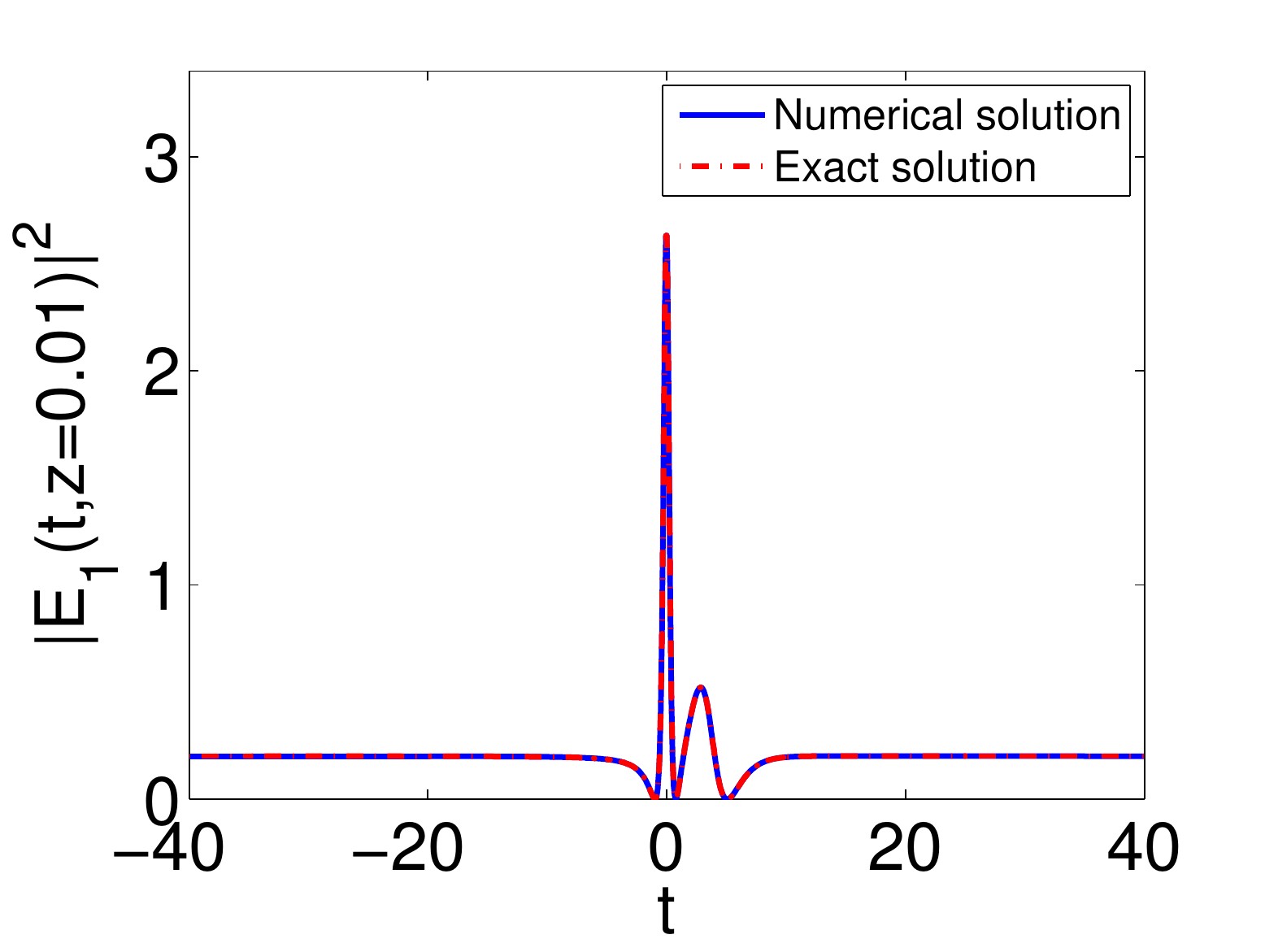}
\label{fig6a}
}
\subfigure[][]{\hspace{-0.2cm}
\includegraphics[height=.21\textheight, angle =0]{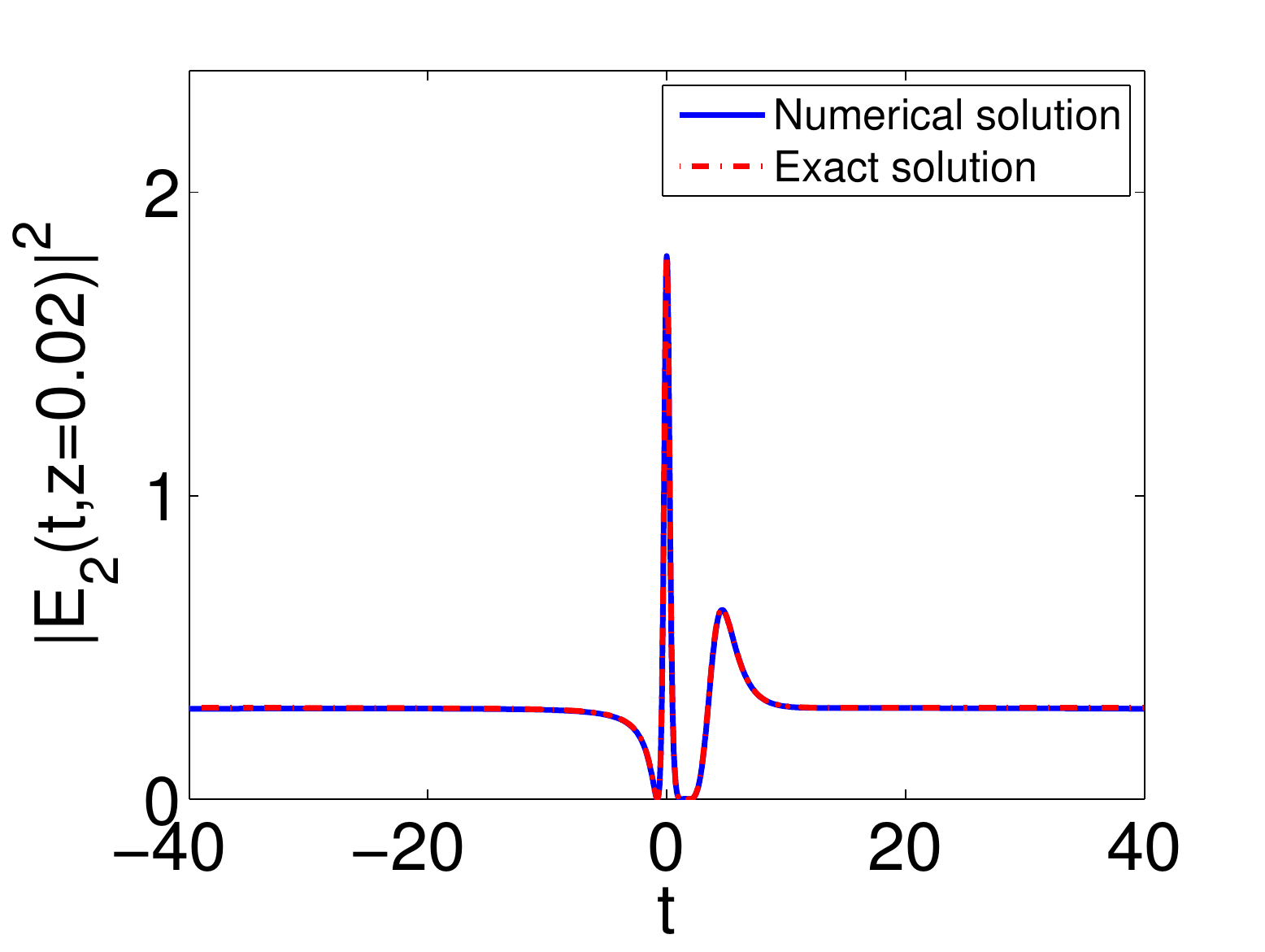}
\label{fig6b}
}
}
\mbox{\hspace{-0.2cm}
\subfigure[][]{\hspace{-1.0cm}
\includegraphics[height=.21\textheight, angle =0]{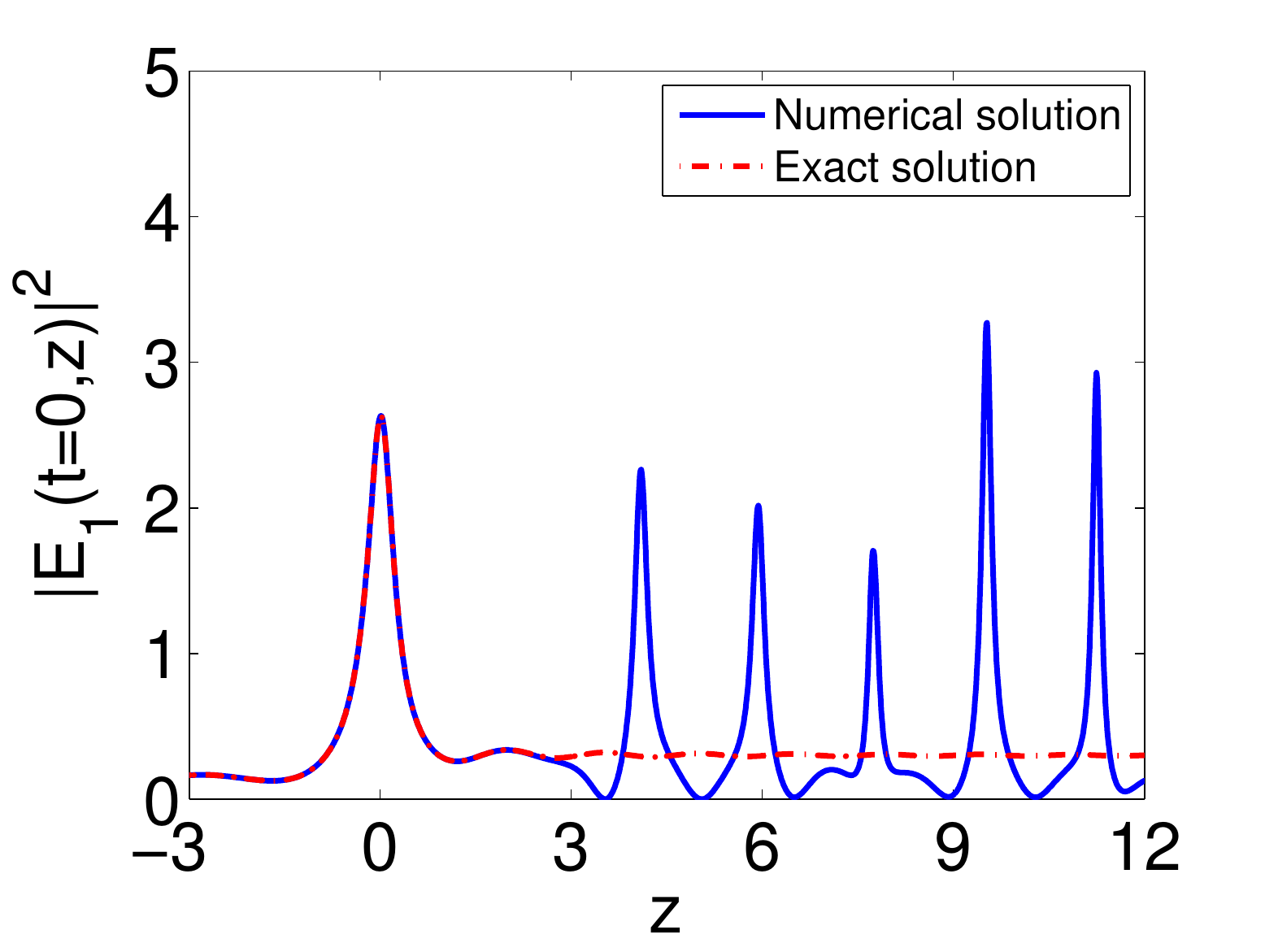}
\label{fig6c}
}
\subfigure[][]{\hspace{-0.2cm}
\includegraphics[height=.21\textheight, angle =0]{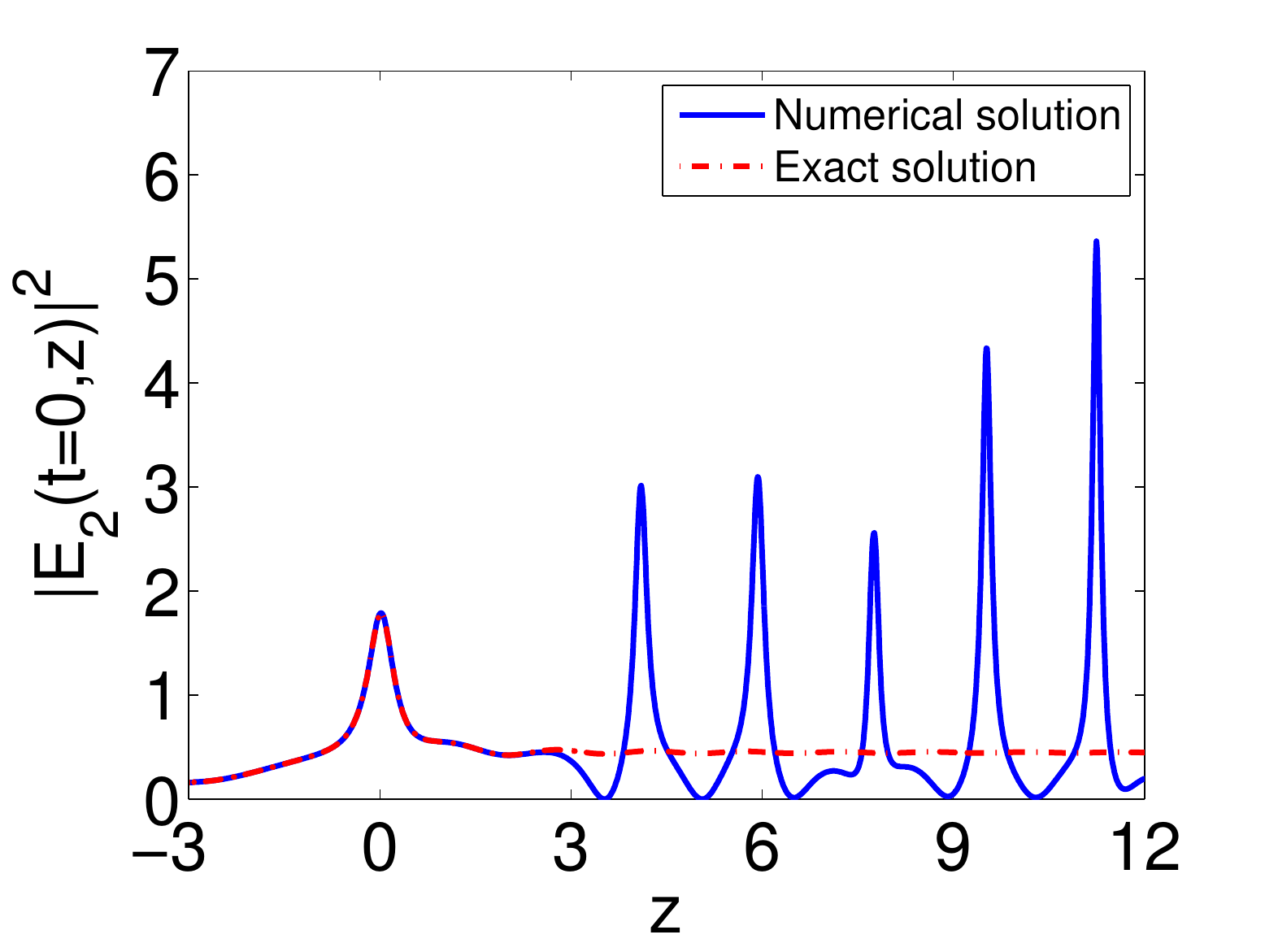}
\label{fig6d}
}
}
\mbox{\hspace{-0.2cm}
\subfigure[][]{\hspace{-1.0cm}
\includegraphics[height=.21\textheight, angle =0]{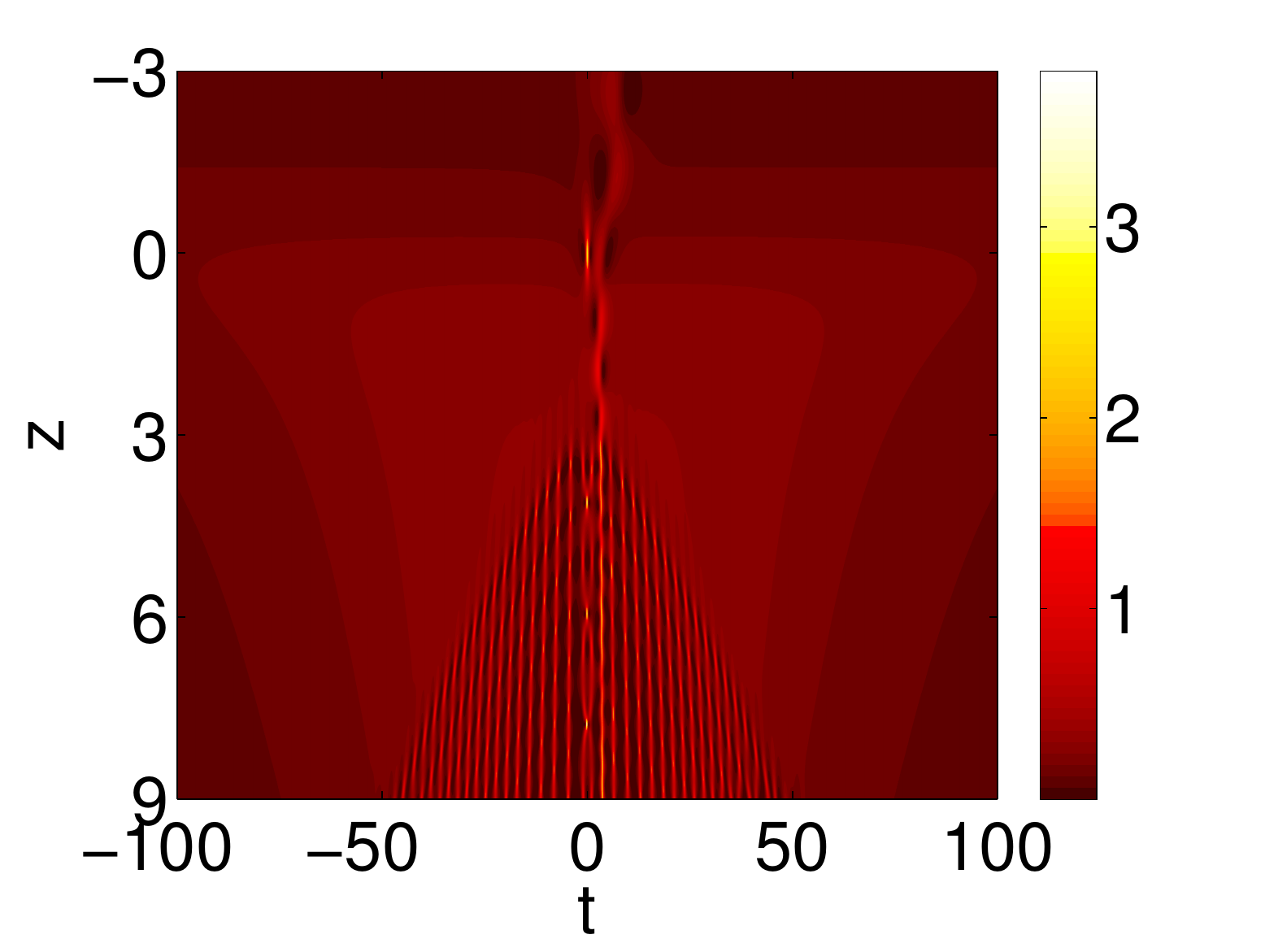}
\label{fig6e}
}
\subfigure[][]{\hspace{-0.2cm}
\includegraphics[height=.21\textheight, angle =0]{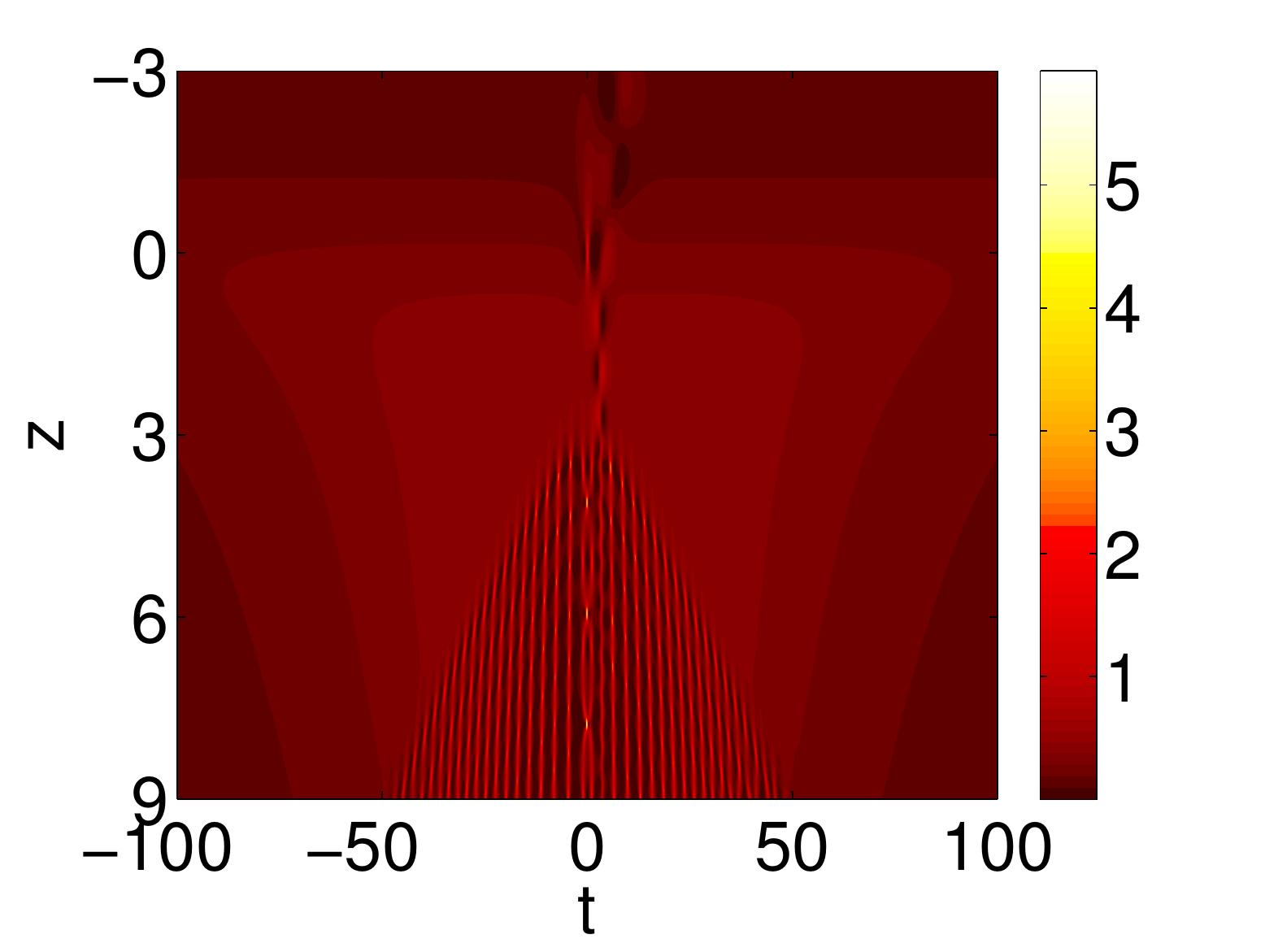}
\label{fig6f}
}
}
\mbox{\hspace{-0.2cm}
\subfigure[][]{\hspace{-1.0cm}
\includegraphics[height=.22\textheight, angle =0]{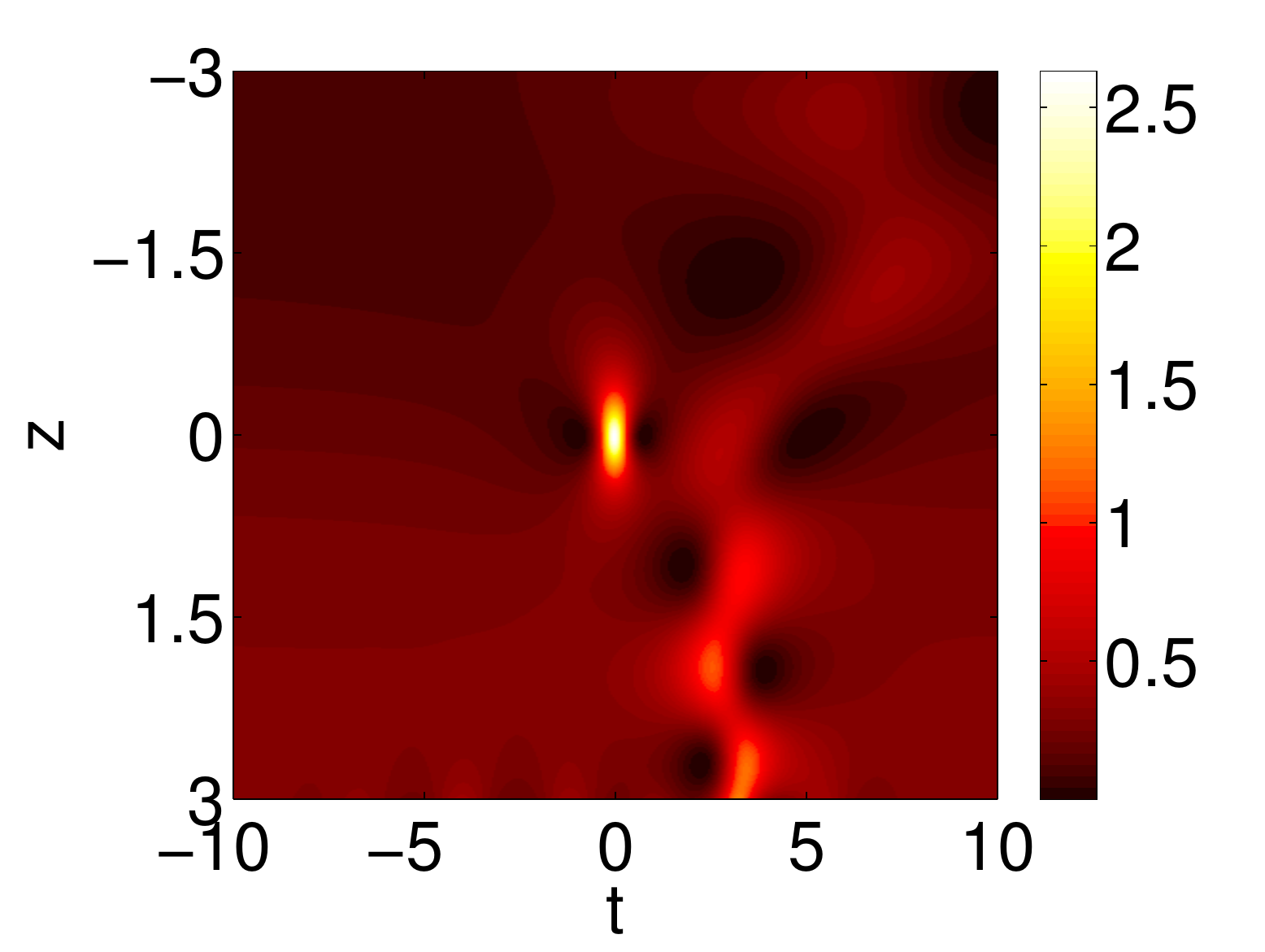}
\label{fig6g}
}
\subfigure[][]{\hspace{-0.2cm}
\includegraphics[height=.22\textheight, angle =0]{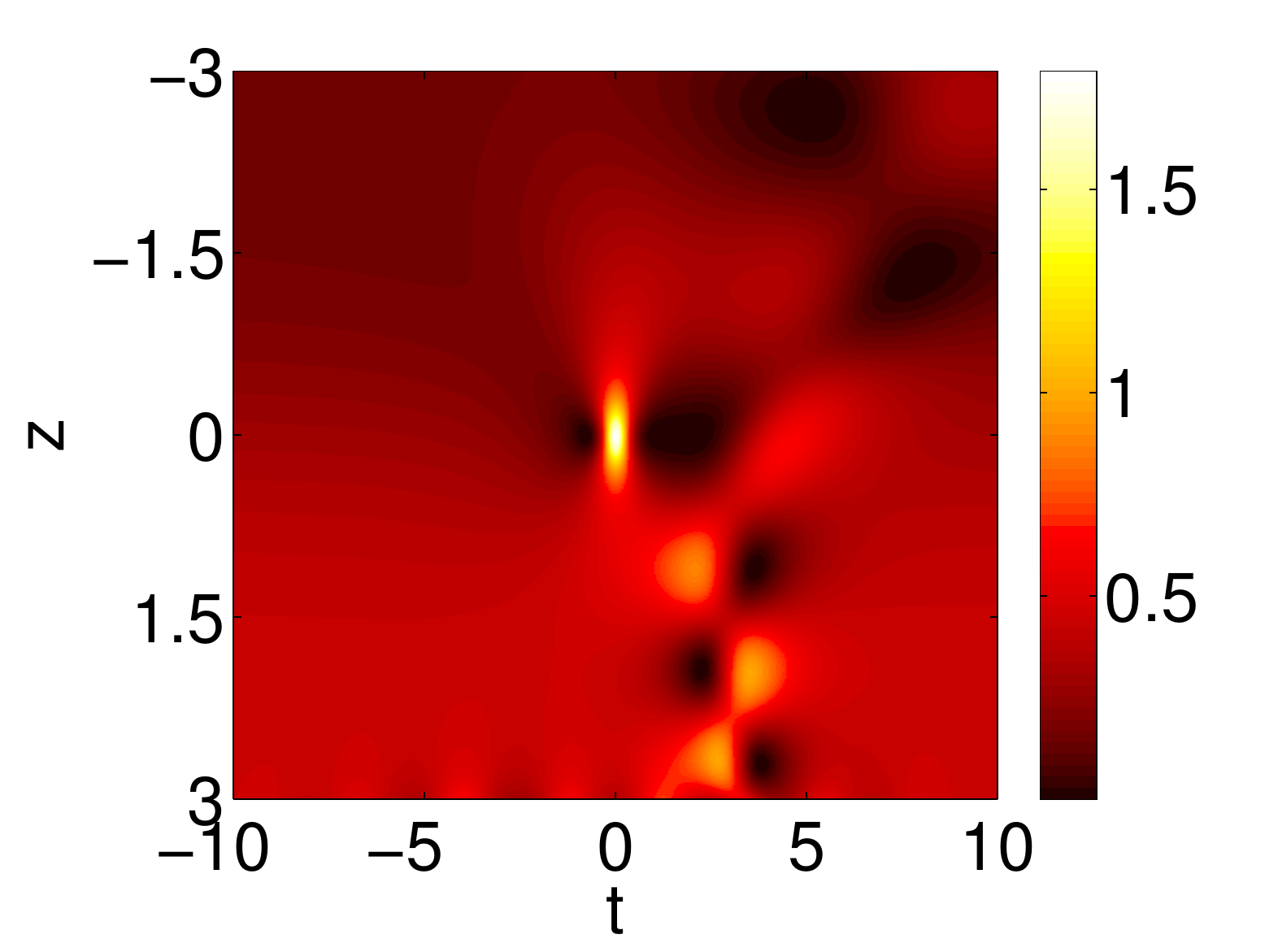}
\label{fig6h}
}
}
\end{center}
\caption{(Color online) Same as Fig.~\ref{fig4} but for the kink-like modulated nonlinearity
given by Eq.~(\ref{eq12a}). Left and right panels correspond to density profiles of the numerically
obtained fields $E_{1}$ and $E_{2}$, respectively. The top row panels (a-b) show the spatial 
distribution of the intensities $|E_{1}|^{2}$ at $z=0.01$ (left) and $|E_{2}|^{2}$ at $z=0.02$
(right), whereas the second row panels (c-d) show their corresponding temporal evolution at $t=0$.
The panels (e-h) show contour plots of the density profiles of the corresponding rogue waves.}
\label{fig6}
\end{figure}

Next, we present results of our numerical simulations for the vector rogue wave solutions of Eqs.~(\ref{inhomogeneous}).
Specifically, Figs.~\ref{fig3} and \ref{fig4} correspond to rogue wave solutions using the periodic
nonlinearity (\ref{non-coefficient}), whereas Figs.~\ref{fig5} and \ref{fig6} correspond to rogue 
wave solutions using the kink-like nonlinearity (\ref{eq12a}). The corresponding final times of the
duration of the simulations reported in this section are $z_{f}=18, 18, 6, 12$, respectively, in these
figures. In addition, the left and right panels correspond to density profiles of the fields $E_{1}$
and $E_{2}$, respectively. Note that panels (a)-(d) demonstrate the density profiles $|E_{j}(t,z)|^2$ ($j=1,2$)
obtained numerically (blue line), in contrast to the corresponding ones from the exact solutions 
(dash-dotted red line). The first row presents the square modulus of the profile of the two fields
as a function of the spatial variable $t$ (at the time-frame of maximum rogue wave density), while the
second one the evolution over $z$ (at the position of maximum rogue wave density).
Finally, panels (e)-(h) depict space-time contour plots of the density profiles of each component 
obtained numerically, and, in particular, panels (g) and (h) of the last row illustrate the local 
densities of each component near the instance of the formation of the rogue wave (near that 
instance, the reversal of direction of the DB boomeronic structure is clearly evident).

\begin{table}[pht]
\caption{Summary of particular cases studied together with parameter values chosen.}
\centering
\begin{tabular}{ | c | c | c | c | c | c | c | c | c | c | }
\hline
 Figure    & Nonlinearity & $a_{1}$ & $a_{2}$ & $\zeta_{1}$ & $\zeta_{2}$ &  $f$   & $\varepsilon$ & $\sigma$ & $\alpha$ \\ \hhline{|==========|}
\ref{fig3} & \multirow{2}{*}{Periodic}   &  $0.8$  &  $0.2$  &     $1$     &     $1$     &  $0$   &    $0.3$   &    $1$   &    $0$   \\ \hhline{-~--------}
\ref{fig4} &              &   $0$   &  $0.3$  &    $1.5$    &     $0$     & $0.15$ &    $0.3$   &   $1.5$  &    $3$   \\ \hline
\ref{fig5} & \multirow{2}{*}{Kink-like}  &  $0.6$  &  $0.8$  &    $0.6$    &     $0$     &  $0$   &    $1.3$   &   $0.55$ &    $2$   \\ \hhline{-~--------}
\ref{fig6} &              &   $0$   &  $0.5$  &     $1$     &     $0$     & $0.15$ &    $0.8$   &    $1$   &   $0.2$   \\
\hline
\end{tabular}
\label{cnls_numer_summary}
\end{table}

Let us now briefly summarize the numerical results presented in this section. It can be discerned immediately
in all the cases studied that vector rogue waves can
be formed (see Figs.~\ref{fig3} and \ref{fig5});
additionally, waveforms involving the co-existence
of rogue waves and DB boomeronic solitons can also naturally
arise from suitable initial data (see Figs.~\ref{fig4} and
\ref{fig6}) in the transformed system.
These solutions directly emerge from the corresponding rogue and
rogue-DB solutions of the regular Manakov problem~\cite{konotop,degasPRL},
yet they bear additional features. In particular, Figs.~\ref{fig3}
and~\ref{fig4} contain a periodic modulation of the profile in $z$
(for a given $t$), while Figs.~\ref{fig5}
and~\ref{fig6} feature a ``jump'' in the field value as the nonlinearity ``step'' is traversed.

However, there is an additional important feature arising
past the formation of these high amplitude solitons. In particular,
the numerically obtained solutions start to
differ from their exact counterparts (except
for the case of Fig.~\ref{fig4} where both exact and numerical solutions agree within the reported
time of the particular simulation). Following the reasoning of the work~\cite{konotop}
(see also~\cite{He2014577}), it can be argued that,
indeed, any destruction of the exact solutions originating from numerical
(e.g., round-off and truncation) errors, which in turn,
contribute to perturbations of the solutions, stems from the
emergence of the
modulational instability (MI)~\cite{kiag}. For instance, in Figs.~\ref{fig3c} and \ref{fig3d}
the numerical solutions, even beyond the
time of occurrence of the high amplitude solitons, accurately capture their exact counterparts, although
at later times deviations arise and eventually lead to progressively increasing
solitary wave patterns. In fact, the same picture holds in Figs.~\ref{fig5c} and \ref{fig5d}, however in this case
the numerical errors have been amplified earlier and past the formation of the vector rogue waves,
the
MI leads to the formation of a soliton train. It perhaps comes as no surprise that the instability's origin is exactly at the very location of the large amplitude waves (such as the rogue waves), as it is in these spots that the rapid growth of the field leads to the most significant amplification of the approximation errors.
It is for that reason that the emerging soliton train arises in the form of a ``sprinkler'' (see, e.g., also~\cite{kody} for a similar
example).

Finally, in Figs.~\ref{fig4} and \ref{fig6} the interaction between
rogue waves and solitons with boomeronic behavior is presented while $f$, $a_{2}\neq0$, and $a_{1}=0$ (see,
for details, Table~\ref{cnls_numer_summary}). According to \cite{degasPRL}, this is the case for the integrable
Manakov system where all the parameters $a_{1}$, $a_{2}$, and $f$ are strictly non-vanishing, and we
expect the emergence of the combined rogue wave-boomeronic DB wave structure.
In Fig.~\ref{fig6}, shortly after
the occurrence of the boomeron-like soliton, the perturbations eventually lead again to MI.
Interestingly, in contrast, this is not the case in Fig.~\ref{fig4} where the
numerical solutions faithfully reproduce the
corresponding exact ones within the time window of the particular simulation. Indeed, the relatively small
amplitude of the field $E_{1}$ compared to its counterpart $E_{2}$,
possibly inhibits any substantial amplification of
numerically induced perturbations for the propagation intervals monitored
herein.
Nevertheless, and despite the avoidance of the manifestation of the
instability in this example, in the vast majority of cases that we
have examined (beyond the typical ones reported here), the instability
has been observed, especially for sizable amplitudes of the
background field, reflecting its significance in terms of the observable
behavior in the system (e.g., when monitored in numerical or physical
experiments).


\section{Conclusions and Future Challenges}
In the present work, we have revisited the theme of settings presenting
modulations in the evolution variable by combining a few relevant, tractable
features therein. On the one hand, our starting point was a vector
nonlinear Schr{\"o}dinger system of relevance to both atomic
BECs and nonlinear optics, by virtue of specialized yet experimentally
tractable settings that were proposed. These featured periodic or hyperbolic dependencies on the evolution
variable of the nonlinearity coefficient and the parabolic potential
prefactor. 
We used
suitable transformations to convert the non-autonomous version
of these multi-component systems into their autonomous siblings. Finally,
we explored an intriguing set of recently proposed waveforms
in such systems, namely the rogue wave in the form of a Peregrine
multi-component soliton, as well as a combination of the rogue
wave with a dark-bright solitonic boomeron. Our direct numerical simulations
indicated that, indeed, such waveforms were supported and can be observed
in the modulated, non-autonomous systems, yet their evolution as perturbed by minimal
errors (including round-off and local truncation errors) leads typically
to the prominent manifestation of the modulational instability, which,
in turn, led to the generation of solitonic trains ``emanating''
from the location of the principal, large amplitude waves.

There are numerous future directions that can be envisioned as
the continuation of the present work. On the one hand, while
the current setting refers to ``pseudo-spinor'' systems (i.e.,
bearing two components), genuine spinor BECs
with more than two components are an intense object of study in the realm
of BECs~\cite{ueda}. On the other hand, one can envision generalizing
the present considerations (and transformations) to other
classes of equations, including to two-dimensional generalizations
and non-autonomous variants of models, such as the Kadomtsev-Petviashvili
or the Davey-Stewartson equation among others~\cite{infeld}. Such
systems are currently under active consideration and will be
reported in future publications.
\vspace{0.5cm}

{\bf Acknowledgments.}
%
E.G.C. gratefully acknowledges financial support from the FP7 People
IRSES-606096: ``{\it Topological Solitons, from Field Theory to Cosmos}''. He also thanks
Hans Johnston (UMass) for providing computing facilities. P.G.K. also acknowledges support
from the National Science Foundation under grants CMMI-1000337, DMS-1312856, from the
Binational Science Foundation under grant 2010239, from FP7-People under grant IRSES-606096
and from the US-AFOSR under grant FA9550-12-10332. The work of D.J.F. was partially supported
by the Special Account for Research Grants of the University of Athens. Finally, the work of 
T.K. and R.B.M. was partially supported by the DST, India under Grant No.SR/S2/HEP-19/2009.

\end{document}